\RequirePackage{etoolbox}
\csdef{input@path}{%
 {sty/}
 {img/}
}%
\csgdef{bibdir}{bib/}

\documentclass[ba]{imsart}
\pubyear{2015}
\volume{10}
\issue{1}
\doi{10.1214/14-BA889}
\firstpage{109}
\lastpage{138}

\usepackage{amsmath, amssymb, amsfonts}
\usepackage{booktabs}
\usepackage{multirow}
\usepackage{algorithm}

\usepackage{amsthm}
\usepackage{natbib}
\usepackage[colorlinks,citecolor=blue,urlcolor=blue,filecolor=blue,backref=page]{hyperref}
\usepackage{graphicx}
\usepackage{subfigure}

\newtheorem{theorem}{Theorem}[section]

\newcommand{\Ehat}{\overline{E}}
\newcommand{\xb}{\mathbf{x}}

\begin{document}

\begin{frontmatter}

\title{Bayesian Structure Learning in Sparse Gaussian Graphical Models}
\runtitle{ Bayesian Structure Learning in Graphical Models }

\begin{aug}
\author{ 
\fnms{Reza} \snm{Mohammadi} \thanksref{addr1} \ead[label=e1]{a.mohammadi@uva.nl} 
\ead[label=e2,url]{http://www.uva.nl/profile/a.mohammadi}
and
\fnms{Ernst C.}  \snm{Wit} \thanksref{addr2} \ead[label=e3]{e.c.wit@rug.nl} 
}
\runauthor{R. Mohammadi and E. C. Wit}
\address[addr1]{ Dept. of Operation Management, University of Amstrdam, Netherlands, \printead{e1},  \printead{e2}}
\address[addr2]{ Institute of Computational Science, Universita della Svizzera Italiana, Lugano, Switzerland, \printead{e3}}
\end{aug}

\begin{abstract}
  Decoding complex relationships among large numbers of variables with relatively few observations is one of the crucial issues in science. One approach to this problem is Gaussian graphical modeling, which describes conditional independence of variables through the presence or absence of edges in the underlying graph. In this paper, we introduce a novel and efficient Bayesian framework for Gaussian graphical model determination  which is a trans-dimensional Markov Chain Monte Carlo (MCMC) approach based on a continuous-time birth-death process. We cover the theory and computational details of the method. It is easy to implement and computationally feasible for high-dimensional graphs. We show our method outperforms alternative Bayesian approaches in terms of convergence, mixing in the graph space and computing time. Unlike frequentist approaches, it gives a principled and, in practice, sensible approach for structure learning. We illustrate the efficiency of the method on a broad range of simulated data. We then apply the method on large-scale real applications from human and mammary gland gene expression studies to show its empirical usefulness. In addition, we implemented the method in the R package {\tt BDgraph} which is freely available at \url{http://CRAN.R-project.org/package=BDgraph}.
\end{abstract}


\end{frontmatter}

\section[sec:intro]{Introduction}

Statistical inference of complex relationships among large numbers of variables with a relatively small number of observations appears in many circumstances. Biologists want to recover the underlying genomic network between thousands of genes, based on at most a few hundred observations. In market basket analysis analysts try to find relationships between only a small number of purchases of individual customers \citep{giudici2003improving}. One approach to these tasks is probabilistic graphical modeling \citep{lauritzen1996graphical}, which is based on the conditional independencies between variables. Graphical models offer fundamental tools to describe the underlying conditional correlation structure. They have recently gained in popularity in both statistics and machine learning with the rise of high-dimensional data \citep{jones2005experiments, dobra2011bayesian, meinshausen2006high, wang2012efficient, friedman2008sparse, ravikumar2010high, zhao2006model, wang2012bayesian, wangscaling}. For the purpose of performing structure learning, Bayesian approaches provide a straightforward tool, explicitly incorporating underlying graph uncertainty.

In this paper, we focus on Bayesian structure learning in Gaussian graphical models for both decomposable and non-decomposable cases. Gaussian graphical determination can be viewed as a covariance selection problem \citep{dempster1972covariance}, where the non-zero entries in the off-diagonal of the precision matrix correspond to the edges in the graph. For a $p$-dimensional variable there are in total $2^{p(p-1)/2}$ possible conditional independence graphs. Even with a moderate number of variables, the model space is astronomical in size. The methodological problem as the dimension grows includes searching over the graph space to identify high posterior regions. High-dimensional regimes, such as genetic networks, have hundreds of nodes, resulting in over $10 ^ {100}$ possible graphs. This motivates us to construct an efficient search algorithm which explores the graph space to distinguish important edges from irrelevant ones and detect the underlying graph with high accuracy. One solution is the trans-dimensional MCMC methodology \citep{green2003trans}. 

In the trans-dimensional MCMC methodology, the MCMC algorithm explores the model space to identify high posterior probability models and estimate the parameters simultaneously. A special case is the reversible-jump MCMC (RJMCMC) approach, proposed by \citet{green1995reversible}. This method constructs an ergodic discrete-time Markov chain whose stationary distribution is taken to be the joint posterior distribution of the model and the parameters. The process transits among models using an acceptance probability, which guarantees convergence to the target posterior distribution. If this probability is high, the process efficiently explores the model space. However, for the high-dimensional regime this is not always efficient. \citet{giudici1999decomposable} extended this method for the decomposable Gaussian graphical models. \citet{dobra2011bayesian} developed it based on the Cholesky decomposition of the precision matrix. \citet{lenkoski2013direct}, \citet{wang2012efficient}, and \citet{cheng2012hierarchical} developed an RJMCMC algorithm, which combined the exchange algorithm \citep{murray2012mcmc} and the double Metropolis-Hastings algorithm \citep{liang2010double} to avoid the intractable normalizing constant calculation.

An alternative trans-dimensional MCMC methodology is the birth-death MCMC (BDMCMC) approach, which is based on a continuous time Markov process. In this method, the time between jumps to a larger dimension (birth) or a smaller one (death) is taken to be a random variable with a specific rate. The choice of birth and death rates determines the birth-death process, and is made in such a way that the stationary distribution is precisely the posterior distribution of interest. Contrary to the RJMCMC approach, moves between models are always accepted, which makes the BDMCMC approach extremely efficient.  In the context of finite mixture distributions with variable dimension this method has been used \citep{stephens2000bayesian}, following earlier proposals by \citet{ripley1977modelling} and \citet{geyer1994simulation}. 
More recently, this type of algorithm, has been used \citep{dobra2018loglinear, mohammadi2017ratio, mohammadi2016bayesian} in the context of graphical models and has been used \citep{mohammadi2019conituous} in the context of Bayesian regression tree.

The main contribution of this paper is to introduce a novel Bayesian framework for Gaussian graphical model determination and design a BDMCMC algorithm to perform both structure learning (graph estimation) and parameter learning (parameters estimation). In our BDMCMC method, we add or remove edges via birth or death events. The birth and death events are modeled as independent Poisson processes. Therefore, the time between two successive birth or death events has an exponential distribution. The birth and death events occur in continuous time and the relative rates at which they occur determine the stationary distribution of the process. The relationships between these rates and the stationary distribution is formalized in Section \ref{sec:BDMCMC} (Theorem \ref{BD theorem}).

The outline of this paper is as follows. In Section \ref{sec:GGM}, we introduce the notation and preliminary background material such as suitable prior distributions for the graph and precision matrix. In Section \ref{sec:BDMCMC}, we propose our Bayesian framework and design our BDMCMC algorithm. In addition, this section contains the specific implementation of our method, including an efficient way for computing the birth and death rates of our algorithm and a direct sampler algorithm from G-Wishart distribution for the precision matrix. In Section \ref{examples}, we show the performance of the proposed method in several comprehensive simulation studies and large-scale real-world examples from human gene expression data and a mouse mammary gland microarray experiment.

\section{Bayesian Gaussian graphical models}
\label{sec:GGM}
 
We introduce some notation and the structure of undirected Gaussian graphical models; for a comprehensive introduction see \citet{lauritzen1996graphical}. Let $G=(V,E)$ be an undirected graph, where $V=\left\{1,2,...,p \right\}$ is the set of nodes and $E \subset V\times V$ is the set of existing edges. Let
\begin{eqnarray*}
\mathcal{W} = \left\{ (i,j) \ | \ i,j \in V, \ i < j \right\},
\end{eqnarray*}
and $\Ehat = \mathcal{W} \backslash E$ denotes the set of non-existing edges. We define a zero mean Gaussian graphical model with respect to the graph $G$ as
\begin{eqnarray*}
\mathcal{M}_G = \left\{ \mathcal{N}_p (0,\Sigma) \ | \ K=\Sigma^{-1} \in \mathbb{P}_{G} \right\},
\end{eqnarray*}
where $\mathbb{P}_{G}$ denotes the space of $p\times p$ positive definite matrices with entries $(i,j)$ equal to zero whenever $(i,j) \in \Ehat$. Let $ \mathbf{x} = (\mathbf{x}^{1},...,\mathbf{x}^{n}) $ be an independent and identically distributed sample of size $n$ from model $\mathcal{M}_G$. Then, the likelihood is
\begin{eqnarray}
\label{likelihood}
P( \mathbf{x} | K,G) \propto |K|^{n/2} \exp \left\{ -\frac{1}{2} \mbox{tr}(KS) \right\},
\end{eqnarray}
where $ S = \mathbf{x}' \mathbf{x}$.

The joint posterior distribution is given as
\begin{eqnarray}
\label{posterior}
P(G,K|\xb) \propto P(\mathbf{x} | G,K) P(K|G) P(G).      
\end{eqnarray}
For the prior distribution of the graph there are many options, of which we propose two. In the absence of any prior beliefs related to the graph structure, one case is a discrete uniform distribution over the graph space $\mathcal{G}$,
\begin{eqnarray*}
P(G)=\frac{1}{|\mathcal{G}|}, \quad \mbox{for each} \quad G \in \mathcal{G}.
\end{eqnarray*}
Alternatively, we propose a truncated Poisson distribution on the graph size $|E|$ with parameter $\gamma$,
\begin{eqnarray*}
p(G) \propto \frac{\gamma^{|E|}}{|E|!}, \quad \mbox{for each} \quad G=(V,E) \in \mathcal{G}.
\end{eqnarray*}
Other choices of priors for the graph structure involve modelling the joint state of the edges as a multivariate discrete distribution \citep{carvalho2009objective} and \citep{scutari2013prior}, encouraging sparse graphs \citep{jones2005experiments} or having multiple testing correction properties \citep{scott2006exploration}.

For the prior distribution of the precision matrix, we use the G-Wishart \citep{roverato2002hyper, letac2007wishart}, which is attractive since it represents the conjugate prior for normally distributed data. It places no probability mass on zero entries of the precision matrix. A zero-constrained random matrix $K \in \mathbb{P}_G$ has the G-Wishart distribution $W_G(b,D)$, if 
\begin{eqnarray*}
P(K|G)=\frac{1}{I_G (b,D)} |K|^{(b-2)/2} \exp \left\{ -\frac{1}{2} \mbox{tr}(DK) \right\},
\end{eqnarray*}
where $b > 2$ is the degree of freedom, $D$ is a symmetric positive definite matrix, and $I_G (b,D)$ is the normalizing constant,
\begin{eqnarray*}
I_G (b,D) = \int_{\mathbb{P}_{G}} |K|^{(b-2)/2} \exp \left\{ -\frac{1}{2} \mbox{tr}(DK) \right\} dK.
\end{eqnarray*}
When $G$ is complete the G-Wishart distribution $W_G(b,D)$ reduces to the Wishart distribution $W_p(b,D)$, hence, its normalizing constant has an explicit form \citep{muirhead1982aspects}. If $G$ is decomposable, we can explicitly calculate $I_G (b,D)$ \citep{roverato2002hyper}. For non-decomposable graphs, however, $I_G (b,D)$ does not have an explicit form; we can numerically approximate $I_G (b,D)$ by the Monte Carlo method \citep{atay2005monte} or Laplace approximation \citep{lenkoski2011computational}.

The G-Wishart prior is conjugate to the likelihood (\ref{likelihood}), hence, conditional on graph $G$ and observed data $\xb$, the posterior distribution of $K$ is
\begin{eqnarray*}
P(K|\xb,G)=\frac{1}{I_G (b^*,D^*)} |K|^{(b^*-2)/2} \exp \left\{ -\frac{1}{2} \mbox{tr}(D^*K) \right\},
\end{eqnarray*}
where $b^*=b+n$ and $D^*=D+S$, that is, $W_G(b^*,D^*)$.

Other choices of priors for the precision matrix are considered on a class of shrinkage priors \citep{wang2013class} using the graphical lasso approach \citep{wang2012bayesian, wangscaling}. They place constant priors for the nonzero entries of the precision matrix and no probability mass on zero entries.

In the following section, we describe an efficient trans-dimensional MCMC sampler scheme for our joint posterior distribution (\ref{posterior}).


\section{The birth-death MCMC method}
\label{sec:BDMCMC}

Here, we determine a continuous time birth-death Markov process particularly for Gaussian graphical model selection. The process explores over the graph space by adding or removing an edge in a birth or death event. The birth and death rates of edges occur  in continuous time with the rates determined by the stationary distribution of the process.

Suppose the birth-death process at time $t$ is at state $(G, K)$ in which $G = (V,E)$ with precision matrix $K \in \mathbb{P}_{G}$. Let $\Omega = \cup_{\substack{G \in \mathcal{G} \\ K \in \mathbb{P}_{G} }} (G, K)$ where $\mathcal{G}$ denotes the set of all possible graphs. We consider the following continuous time birth-death Markov process on $\Omega$:

{\bf Death:} Each edge $e \in E$ dies independently of the others as a Poisson process with a rate $\delta_{e}(K)$. Thus, the overall death rate is $\delta(K) = \sum_{e \in E}{\delta_{e}(K)}$. If the death of an edge $e=(i,j) \in E$ occurs, then the process jumps to a new state $(G^{-e}, K^{-e})$ in which $G^{-e} = ( V, E \setminus \{ e \})$, and $K^{-e} \in \mathbb{P}_{G^{-e}}$. We assume $K^{-e}$ is equal to matrix $K$ except for the entries in positions $\{(i,j), (j,i), (j,j)\}$. Note we can distinguish $i$ from $j$, since by our definition of an edge $i < j$.

{\bf Birth:} A new edge $e \in \Ehat$ is born independently of the others as a Poisson process with a rate $\beta_{e}(K)$. Thus, the overall birth rate is $\beta(K) = \sum_{e \in \Ehat}{\beta_{e}(K)}$. If the birth of an edge $e=(i,j) \in \Ehat$ occurs, then the process jumps to a new state $(G^{+e}, K^{+e})$ in which $G^{+e}=(V, E \cup \{ e \})$, and $K^{+e} \in \mathbb{P}_{G^{+e}}$. We assume $K^{+e}$ is equal to matrix $K$ except for the entries in positions $\{(i,j), (j,i), (j,j)\}$. 

The birth and death processes are independent Poisson processes. Thus, the time between two successive events is exponentially distributed, with mean $1/(\beta(K)+\delta(K))$. Therefore, the probability of a next birth/death event is 

\begin{eqnarray}
\label{prob.birth}
P(\mbox{birth for edge  } e)= \frac{\beta_{e}(K)}{\beta(K)+\delta(K)}, \qquad \mbox{for each} \ \ e \in \Ehat,
\end{eqnarray}

\begin{eqnarray}
\label{prob.death}
P(\mbox{death for edge  } e)= \frac{\delta_{e}(K)}{\beta(K)+\delta(K)}, \qquad \mbox{for each} \ \ e \in E.
\end{eqnarray}

The following theorem gives a sufficient condition for which the stationary distribution of our birth-death process is precisely the joint posterior distribution of the graph and precision matrix.

\begin{theorem}
\label{BD theorem}
The above birth-death process has stationary distribution $P(K,G|\xb)$, if for each $e \in \mathcal{W}$
\begin{equation}
\delta_{e}(K) P(G,K\setminus(k_{ij},k_{jj})|\xb) = \beta_{e}(K^{-e}) P(G^{-e},K^{-e}\setminus k_{jj}|\xb).
\end{equation}
\end{theorem}

{\bf Proof.} Our proof is based on the theory derived by \citet*[Section 7 and 8]{preston1976}. Preston proposed a special birth-death process, in which the birth and death rates are functions of the state. The process evolves by two types of jumps: a {\it birth} is defined as the appearance of a single individual, whereas a {\it death} is the removal of a single individual. This process converges to a unique stationary distribution, if the balance conditions hold \citep[Theorem 7.1]{preston1976}. We construct our method in such a way that the stationary distribution equals the joint posterior distribution of the graph and the precision matrix. See the appendix for a detailed proof.

\subsection{Proposed BDMCMC algorithm}
\label{sec:BDMCMC algorithm}

Our proposed BDMCMC algorithm is based on a specific choice of birth and death rates that satisfies Theorem \ref{BD theorem}.  Suppose we consider the birth and death rates as

\begin{eqnarray}
\label{birthrate}
\beta_{e}(K) = \min \left\{ 1, \frac{P(G^{+e},K^{+e} \setminus(k_{ij},k_{jj})|\xb)}{P(G,K\setminus k_{jj}|\xb)} \right\}, \ \ \mbox{for each} \ \ e \in \Ehat,
\end{eqnarray}

\begin{eqnarray}
\label{deathrate}
\delta_{e}(K) =  \min \left\{ 1, \frac{P(G^{-e},K^{-e}\setminus k_{jj}|\xb)}{P(G,K\setminus(k_{ij},k_{jj})|\xb)} \right\}, \: \: \qquad \mbox{for each} \ \ e \in E.
\end{eqnarray}
Based on the above rates, we determine our BDMCMC algorithm as below.
\begin{algorithm}
{\bf Algorithm 3.1. BDMCMC algorithm}. Given a graph $G=(V,E)$ with a precision matrix $K$, iterate the following steps:
\begin{description}
 \item[Step 1.] Birth and death process  
  \begin{description}
   \item[1.1.] Calculate the birth rates by (\ref{birthrate}) and $\beta(K)= \sum_{e \in \Ehat}{\beta_{e}(K)}$,
   \item[1.2.] Calculate the death rates by (\ref{deathrate}) and $\delta(K)=\sum_{e \in E}{\delta_{e}(K)}$,
   \item[1.3.] Calculate the waiting time by $w(K)= 1/(\beta(K)+\delta(K))$,
   \item[1.4.] Simulate the type of jump (birth or death) by (\ref{prob.birth}) and (\ref{prob.death}).
  \end{description}
 \item[Step 2.] According to the type of jump, sample from the new precision matrix.
\end{description}
\end{algorithm}

The main computational parts of our BDMCMC algorithm are computing the birth and death rates (steps 1.1 and 1.2) and sampling from the posterior distribution of the precision matrix (step 2). In Section \ref{computing rates}, we illustrate how to calculate the birth and death rates. In Section \ref{Sample gwishart}, we explain a direct sampling algorithm from the G-Wishart distribution for sampling from the posterior distribution of the precision matrix. 

\begin{figure}[!ht]
\centering
\includegraphics[width=0.8\textwidth]{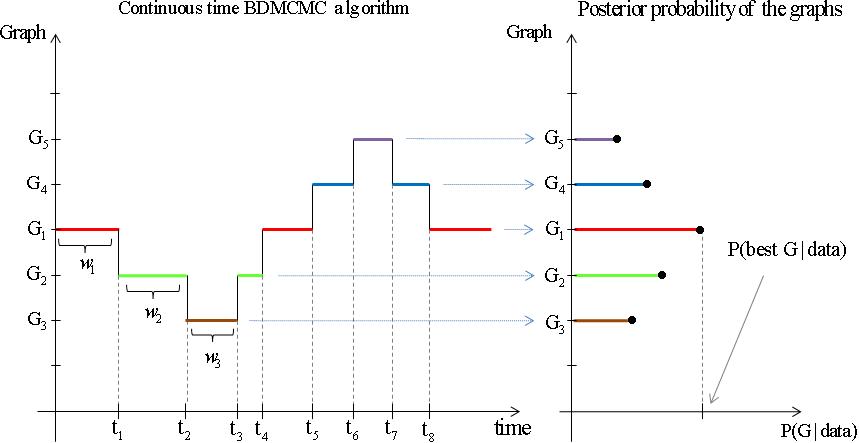}
\caption{ \label{fig:BDMCMC}(Left) Continuous time BDMCMC algorithm where $ \left\{ t_1, t_2, t_3,... \right\}$ are jumping times and $ \left\{ w_1, w_2, w_3,... \right\}$ are waiting times. (Right) Posterior probability estimation of the graphs based on the proportions of their waiting times.}
\end{figure}

In our continuous time BDMCMC algorithm we sample in each step of jumping to the new state (e.g. $ \left\{ t_1, t_2, t_3,... \right\}$ in Figure \ref{fig:BDMCMC}). For inference, we put the weight on each state to effectively compute the sample mean as a Rao-Blackwellized estimator \citep[subsection 2.5]{cappé2003reversible}; See e.g. (\ref{posterior-edge}). The weights are equal to the length of the waiting time in each state ( e.g. $ \left\{ w_1, w_2, w_3,... \right\}$ in Figure \ref{fig:BDMCMC}). Based on these waiting times, we estimate the posterior distribution of the graphs, which are the proportion to the total waiting times of each graph (see Figure \ref{fig:BDMCMC} in the right and Figure \ref{fig:6node-post}). For more detail about sampling from continuous time Markov processes see \citet[subsection 2.5]{cappé2003reversible}.

\subsection{Step1: Computing the birth and death rates}
\label{computing rates}

In step 1 of our BDMCMC algorithm, the main task is calculating the birth and death rates (steps 1.1 and 1.2); Other steps are straightforward. Here, we illustrate how to calculate the death rates. The birth rates are calculated in a similar manner, since both birth and death rates (\ref{birthrate}) and (\ref{deathrate}) are the ratio of the conditional posterior densities. 

For each $e=(i,j) \in E$, the numerator of the death rate is 
\begin{eqnarray*}
P(G^{-e},K^{-e}\setminus k_{jj}|\xb)= \frac{P(G^{-e},K^{-e}|\xb)}{P(k_{jj}|K^{-e}\setminus k_{jj},G^{-e}, \xb)}.
\end{eqnarray*}
The full conditional posterior for $k_{jj}$ is (see \citealt[Lemma 1]{roverato2002hyper})
\begin{eqnarray*}
k_{jj} - c \: | \: K^{-e}\setminus k_{jj},G^{-e}, \xb \sim W(b^*, D^*_{jj}),
\end{eqnarray*}
where $c = K_{j,V \setminus j} (K_{V \setminus j, V \setminus j})^{-1} K_{V \setminus j, j}$. Following \citet{wang2012efficient} and some simplification, we have
\begin{eqnarray}
\label{post-}
P \! (G^{-e} \! ,K^{-e} \! \setminus \! k_{jj}|\xb) \! = \! \frac{P \! (G)}{P \! (\xb)} \frac{I(b^*,D^*_{jj})}{I_{G^{-e}} (b,D)} 
|K^0_{V \setminus j, V \setminus j}|^{(b^* \! - \! 2)/2} \exp \! \left\{ \! -\frac{1}{2} \mbox{tr}(K^0 D^*) \! \right\} \!  \! ,
\end{eqnarray}
where $I(b^*,D^*_{jj})$ is the normalizing constant of a G-Wishart distribution for $p=1$ and $K^0=K$ except for an entry $0$ in the positions $(i,j)$ and $(j,i)$, and an entry $c$ in the position $(j,j)$.

For the denominator of the death rate we have
\begin{eqnarray*}
P(G,K \setminus (k_{ij},k_{jj})|\xb)= \frac{P(G,K|\xb)}{P((k_{ij},k_{jj}) | K \setminus (k_{ij},k_{jj}),G, \xb)},
\end{eqnarray*}
in which we need the full conditional distribution of $(k_{ij},k_{jj})$. We can obtain the conditional posterior of $(k_{ii},k_{ij},k_{jj})$ and by further conditioning on $k_{ii}$ and using the proposition in the appendix, we can evaluate the full conditional distribution of $(k_{ij},k_{jj})$. Following \citet{wang2012efficient} and some simplification, we have
\begin{eqnarray}
\label{post}
P \! ( \! K \! \setminus \! (k_{ij},k_{jj}),G | \xb \! ) \! = \! \frac{ \! P(G)}{P \! (\xb)} \frac{J \! (b^* \! , \! D^*_{e e} \! , \! K)}{I_G(b,D)} 
|K^1_{V \setminus e, V \setminus e}|^{(b^*-2)/2}  \! \exp \! \left\{ \! -\frac{1}{2} \mbox{tr} \! ( \! K^1 D^* \! ) \! \right\} \!  \! ,
\end{eqnarray}
where
\begin{eqnarray*}
J(b^*,D^*_{e e},K) \! = \! (\frac{2 \pi}{D^*_{jj}})^{\frac{1}{2}} I(b^*,D^*_{jj}) (k_{ii}-k^1_{ii})^{\frac{b^* \! - \! 2}{2}} 
                                  \! \exp \! \left\{ \! -\frac{1}{2} (D^*_{ii} - \frac{D^{*2}_{ij}}{D^*_{jj}}) (k_{ii}-k^1_{ii}) \! \right\} \! ,
\end{eqnarray*}
and $K^1=K$ except for the entries $K_{e,V \setminus e} (K_{V \setminus e, V \setminus e})^{-1} K_{V \setminus e, e}$ in the positions corresponding to $e=(i,j)$.

By plugging (\ref{post-}) and (\ref{post}) into the death rates (\ref{deathrate}), we have
\begin{eqnarray}
\label{death rates}
\delta_{e} (K) &=&  \min \left\{ 1, \frac{P(G^{-e})}{P(G)} \frac{I_{G}(b,D)}{I_{G^{-e}}(b,D)} H(K, D^*, e) \right\},  
\end{eqnarray}
in which 
\begin{eqnarray}
\label{Hij}
H \! (K \! , D^* \! , e) \! &=& \! (\frac{D^*_{jj}}{2 \pi(k_{ii}-k^1_{ii})}) ^ {\frac{1}{2}} \nonumber \\
     \! &\times& \! \exp \! \left\{ \! -\frac{1}{2} \! \left[ \! \mbox{tr}(D^{*}(K^0 \! - \! K^1)) - (D^*_{ii} \! - \! 
      \frac{D^{*2}_{ij}}{D^*_{jj}}) (k_{ii} \! - \! k^1_{ii}) \! \right]  \! \right\} \! . 
\end{eqnarray}
For computing the above death rates, we require the prior normalizing constants which is the main computational part. Calculation time for the remaining elements is extremely fast. 

\subsubsection{Coping with evaluation of prior normalizing constants}
\label{computing ratio}


\citet{murray2012mcmc} proved that the exchange algorithm based on exact sampling is a powerful tool for general MCMC algorithms in which their likelihoods have additional parameter-dependent normalization terms, such as the posterior over parameters of an undirected graphical model. \citet{wang2012efficient} and \citet{lenkoski2013direct} illustrate how to use the concept behind the exchange algorithm to circumvent intractable normalizing constants as in (\ref{death rates}). With the existence of a direct sampler of G-Wishart, \citet{lenkoski2013direct} used a modification of the exchange algorithm to cope with the ratio of prior normalizing constants. For an explicit analytic approximation for the ratio of prior normalizing constants see \citet{mohammadi2017ratio}.

Suppose that $(G, K)$  is the current state of our algorithm and we would like to calculate the death rates (\ref{death rates}), first we sample $\tilde{K}$ according to $W_G(b, D)$ via an exact sampler, Algorithm 3.2 below. Then, we replace the death rates with

\begin{eqnarray}
\label{death rates2}
\delta_{e} (K) &=&  \min \left\{ 1, \frac{P(G^{-e})}{P(G)} \frac{H(K, D^*, e)}{H(\tilde{K}, D, e)} \right\}, 
\end{eqnarray}
in which the intractable prior normalizing constants have been replaced by an evaluation of $H$ (given in (\ref{Hij})) in the prior, evaluated at $\tilde{K}$; For theoretical justifications of this procedure, see \citet{murray2012mcmc} and \citet{liang2010double}. 

\subsection{Step 2: Direct sampler from precision matrix}
\label{Sample gwishart}

\citet{lenkoski2013direct} developed an exact sampler method for the precision matrix, which borrows ideas from \citet{hastie2009elements}. The algorithm is as follows.
\begin{algorithm}
 \label{K sampler}
{\bf Algorithm 3.2. Direct sampler from precision matrix} \citet{lenkoski2013direct}. 
Given a graph $G=(V,E)$ with precision matrix $K$ and $\Sigma = K^{-1}$ :
\begin{description}
\item[Step 1.] Set $\Omega = \Sigma$.
\item[Step 2.] Repeat for $i = 1, ..., p$, until convergence:
    \begin{description}
    \item[2.1] Let $N_i \subset V$ be the set of neighbors of node $i$ in graph $G$. Form $\Omega_{N_i}$ and $\Sigma_{N_i,i}$ and solve 
    $$\hat{ \beta_{i}^{*} } = \Omega^{-1}_{N_i} \Sigma_{N_i, i},$$
    \item[2.2] Form $\hat{ \beta_{i} } \in R^{p-1}$ by padding the elements of $\hat{ \beta_{i}^{*} }$ to the appropriate locations and zeroes in those locations not connected to $i$ in graph $G$,
    \item[2.3] Update $\Omega_{i,-i}$ and $\Omega_{-i,i}$ with $\Omega_{-i,-i}\hat{ \beta_{i} }$.
    \end{description}
\item[Step 3.] Return $K = \Omega^{-1}$.
\end{description}
\end{algorithm}

Throughout, we use the direct sampler algorithm for sampling from the precision matrix $K$.

\section{Statistical performance}
\label{examples}

Here we present the results for three comprehensive simulation studies and two applications to real data sets. In Section \ref{simulation study}, we show that our method outperforms alternative Bayesian approaches in terms of convergence, mixing in the graph space and computing time; Moreover, the model selection properties compare favorably with frequentist alternatives. In Section \ref{example: human gene data}, we illustrate our method on a large-scale real data set related to the human gene expression data. In Section \ref{example: mammary data}, we demonstrate the extension of our method to graphical models which involve time series data. It shows how graphs can be useful in modeling real-world problems such as gene expression time course data. We performed all computations with the R package {\tt BDgraph}, \citep{bdgraph, mohammadi2019bdgraph}.

\subsection{Simulation study}
\label{simulation study}

\subsubsection{Graph with 6 nodes}
\label{example 1}

We illustrate the performance of our methodology and compare with two alternative Bayesian methods on a concrete small toy simulation example which comes from \citet{wang2012efficient}. We consider a data generating mechanism with $p=6$ within
\begin{eqnarray*}
\mathcal{M}_G= \left\{ \mathcal{N}_6 (0,\Sigma) \ | \ K=\Sigma^{-1} \in \mathbb{P}_{G} \right\},
\end{eqnarray*}
in which the precision matrix is
\begin{eqnarray*} 
K =
\begin{bmatrix}
 1  & 0.5  & 0   & 0   & 0   & 0.4  \\
    & 1    & 0.5 & 0   & 0   & 0    \\
    &      & 1   & 0.5 & 0   & 0    \\
    &      &     & 1   & 0.5 & 0    \\
    &      &     &     & 1   & 0.5  \\
    &      &     &     &     & 1    \\
\end{bmatrix}.
\end{eqnarray*}
Just like \citet{wang2012efficient} we let $S = n K^{-1}$ where $n=18$, which represents $18$ samples from the true model $\mathcal{M}_G$. As a non-informative prior, we take a uniform distribution for the graph and a G-Wishart $W_G(3,I_6)$ for the precision matrix. 

To evaluate the performance of our BDMCMC algorithm, we run our BDMCMC algorithm with $60,000$ iterations and $30,000$ as a burn-in. All the computations for this example were carried out on an Intel(R) Core(TM) i5 CPU 2.67GHz processor.

We calculate the posterior pairwise edge inclusion probabilities based on the Rao-Blackwellization \citep[subsection 2.5]{cappé2003reversible} as
\begin{eqnarray}
\label{posterior-edge}
\hat{p}_{e}= \frac{\sum_{t=1}^{N}{I(e \in G^{(t)}) w(K^{(t)}) }}{\sum_{t=1}^{N}{w(K^{(t)})}}, \quad \mbox{for each} \quad e \in \mathcal{W},
\end{eqnarray}
where $N$ is the number of iterations, $I(e \in G^{(t)})$ is an indicator function, such that $I(e \in G^{(t)})=1$ if $e \in G^{(t)}$ and zero otherwise, and $w(K^{(t)})$ is the waiting time in the graph $G^{(t)}$ with the precision matrix $K^{(t)}$; see Figure \ref{fig:BDMCMC}. The posterior pairwise edge inclusion probabilities for all the edges $e=(i,j) \in \mathcal{W}$ are
\begin{eqnarray*} 
\hat{p}_e =
\begin{bmatrix}
 0     & 0.98  & 0.05  & 0.02  & 0.03  & 0.92  \\ 
       & 0     & 0.99  & 0.04  & 0.01  & 0.04  \\ 
       &       & 0    & 0.99  & 0.04  & 0.02   \\ 
       &       &        & 0   & 0.99  & 0.06   \\ 
       &       &        &        & 0  & 0.98   \\ 
       &       &        &        &    & 0      \\ 
\end{bmatrix}.
\end{eqnarray*} 
The posterior mean of the precision matrix is
\begin{eqnarray*}  
\hat{K} =
\begin{bmatrix}
1.16  & 0.58  & -0.01   & 0.00   & -0.01   & 0.44    \\
      & 1.18  & 0.58    & -0.01  & 0.00    & -0.01   \\
      &       & 1.18    & 0.58   & -0.01   & 0.00    \\
      &       &         & 1.18   & 0.58    & -0.01   \\
      &       &         &        & 1.17   & 0.57     \\
      &       &         &        &         & 1.16    \\
\end{bmatrix}.
\end{eqnarray*} 

We compare the performance of our BDMCMC algorithm with two recently proposed trans-dimensional MCMC approaches. One is the algorithm proposed by \citet{lenkoski2013direct} which we call ``Lenkoski''. The other is an algorithm proposed by \citet{wang2012efficient} which we call ``WL''. The R code for the WL approach is available at \url{http://r-forge.r-project.org/projects/bgraph/}.

\begin{figure}[ht]
\centering
\subfigure[BDMCMC]{
    \includegraphics[width=.3\textwidth]{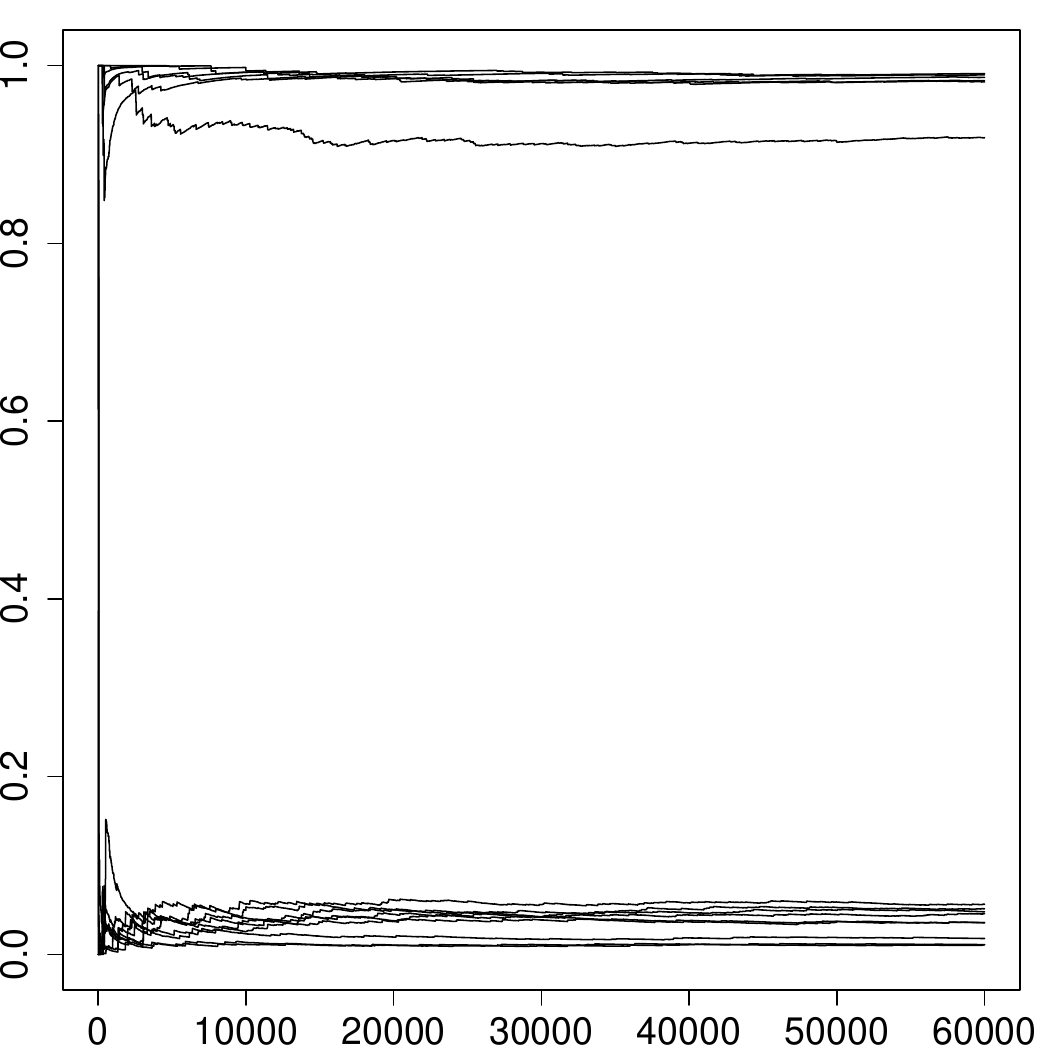}
    \label{fig:bdmcmc}
}
\subfigure[Lenkoski]{
    \includegraphics[width=.3\linewidth]{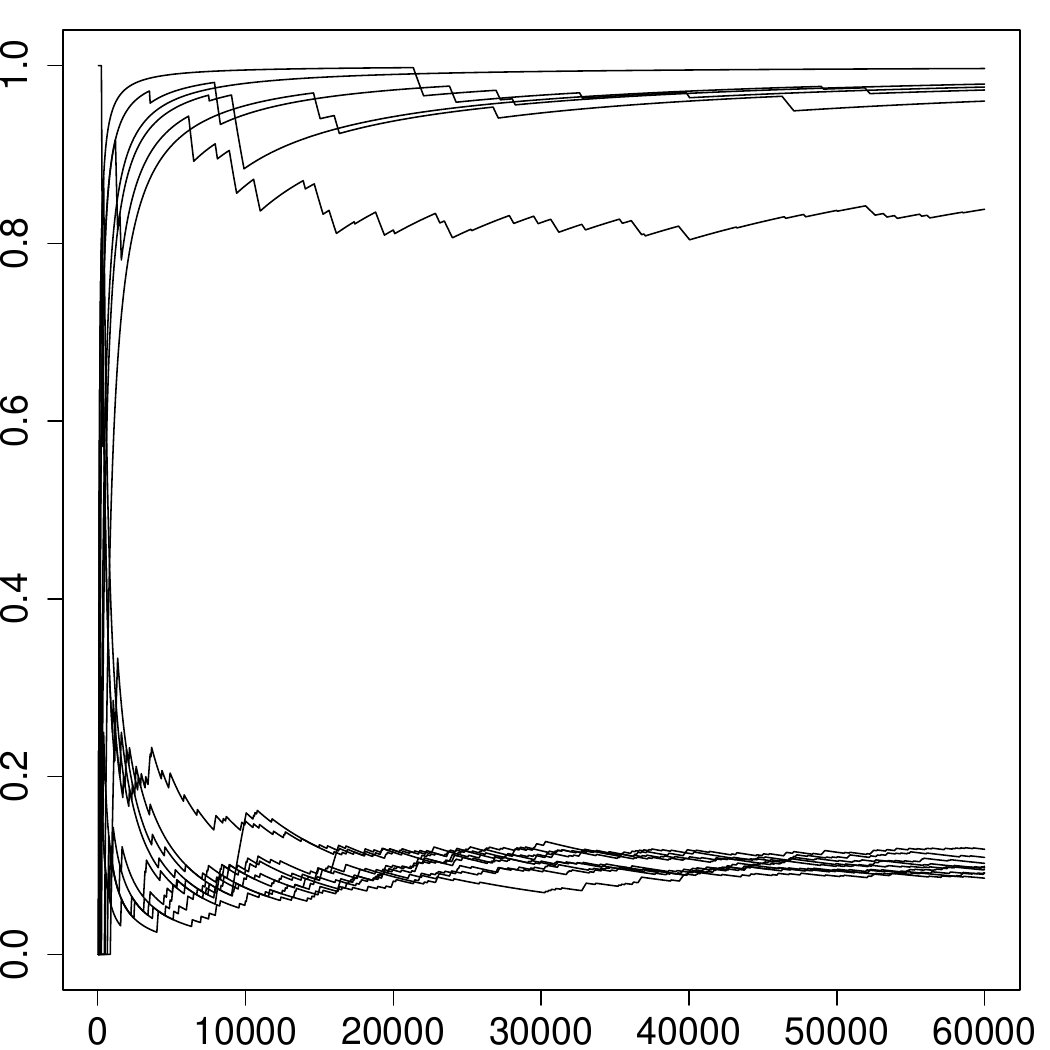}
    \label{fig:drj}
}
\subfigure[WL]{
    \includegraphics[width=.3\textwidth]{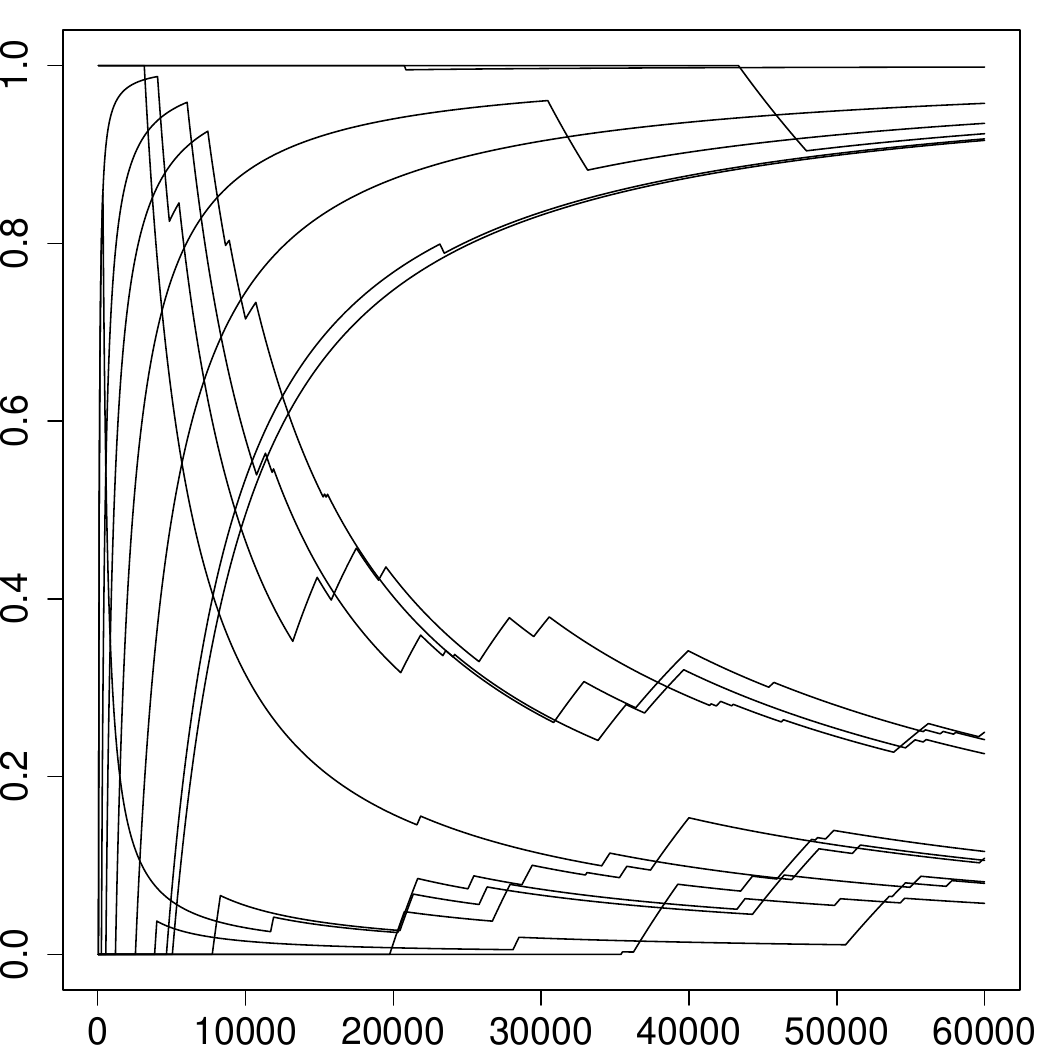}
    \label{fig:dmh}
}
\caption[]{
Plot of the cumulative occupancy fractions of all possible edges to check convergence in simulation example \ref{example 1}; BDMCMC algorithm in \ref{fig:bdmcmc}, Lenkoski algorithm\citep{lenkoski2013direct} in \ref{fig:drj}, and WL algorithm in \citet{wang2012efficient} \ref{fig:dmh}.
}
\label{fig:6node-convergency}
\end{figure}

Compared to other Bayesian approaches, our BDMCMC algorithm is highly efficient due to its fast convergence speed. One useful test of convergence is given by the plot of the cumulative occupancy fraction for all possible edges, shown in Figure \ref{fig:6node-convergency}. Figure \ref{fig:bdmcmc} shows that our BDMCMC algorithm converges after approximately $10,000$ iterations. Figure \ref{fig:drj} and \ref{fig:dmh} show that the Lenkoski algorithm converges after approximately $30,000$, 
whereas the WL algorithm still does not converge after $60,000$ iterations.

\begin{figure}[ht]
\centering
\subfigure[BDMCMC]{
    \includegraphics[width=.3\textwidth]{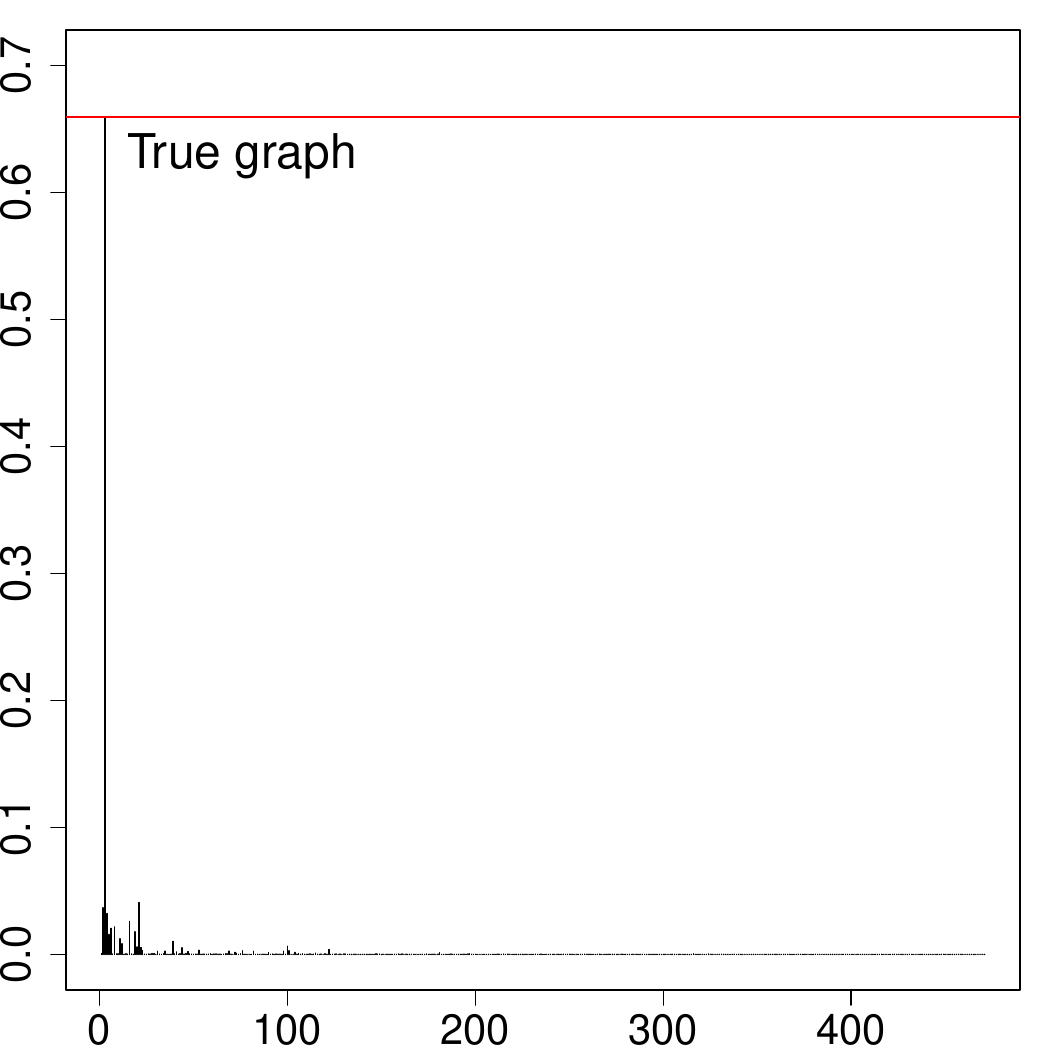}
    \label{fig:6node-post-subfig1}
}
\subfigure[Lenkoski]{
    \includegraphics[width=.3\linewidth]{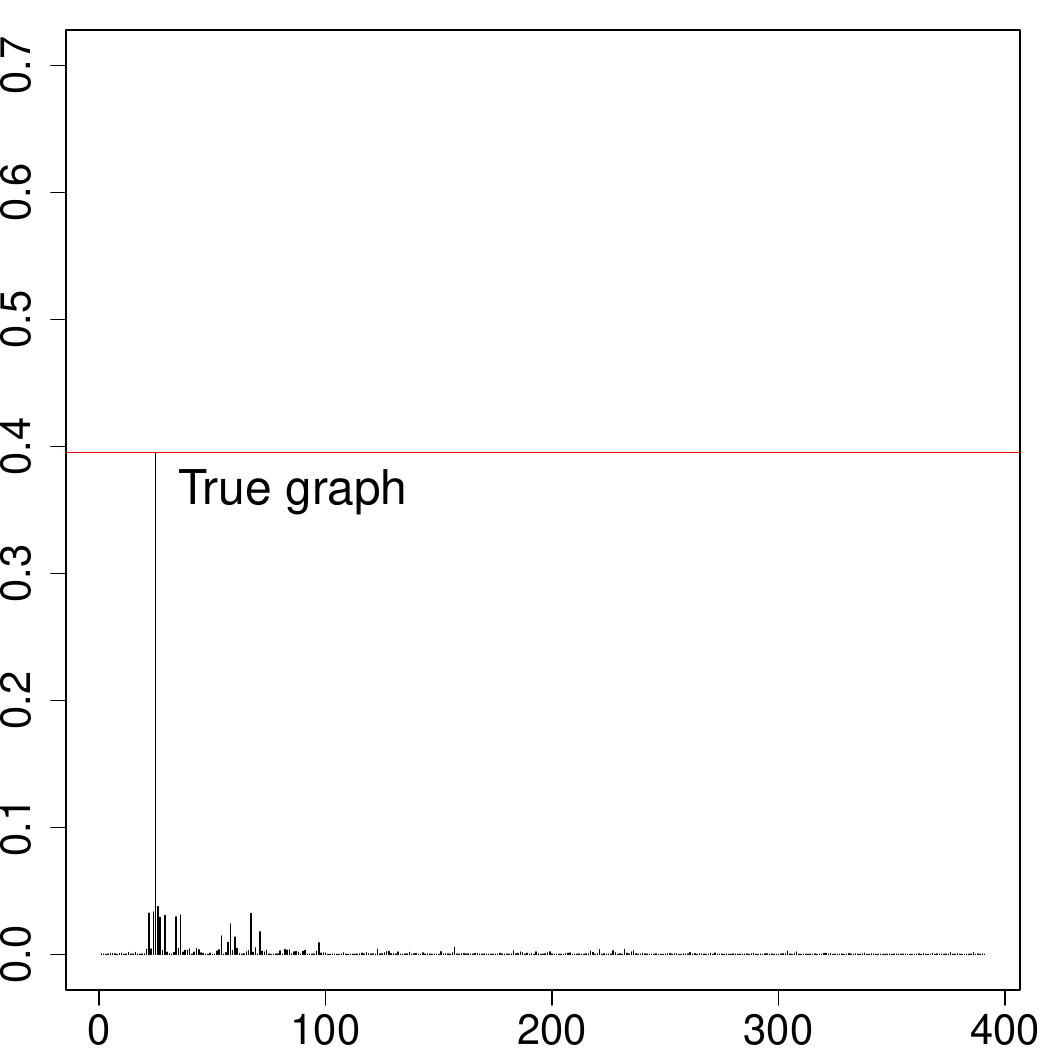}
    \label{fig:6node-post-subfig2}
}
\subfigure[WL]{
    \includegraphics[width=.3\textwidth]{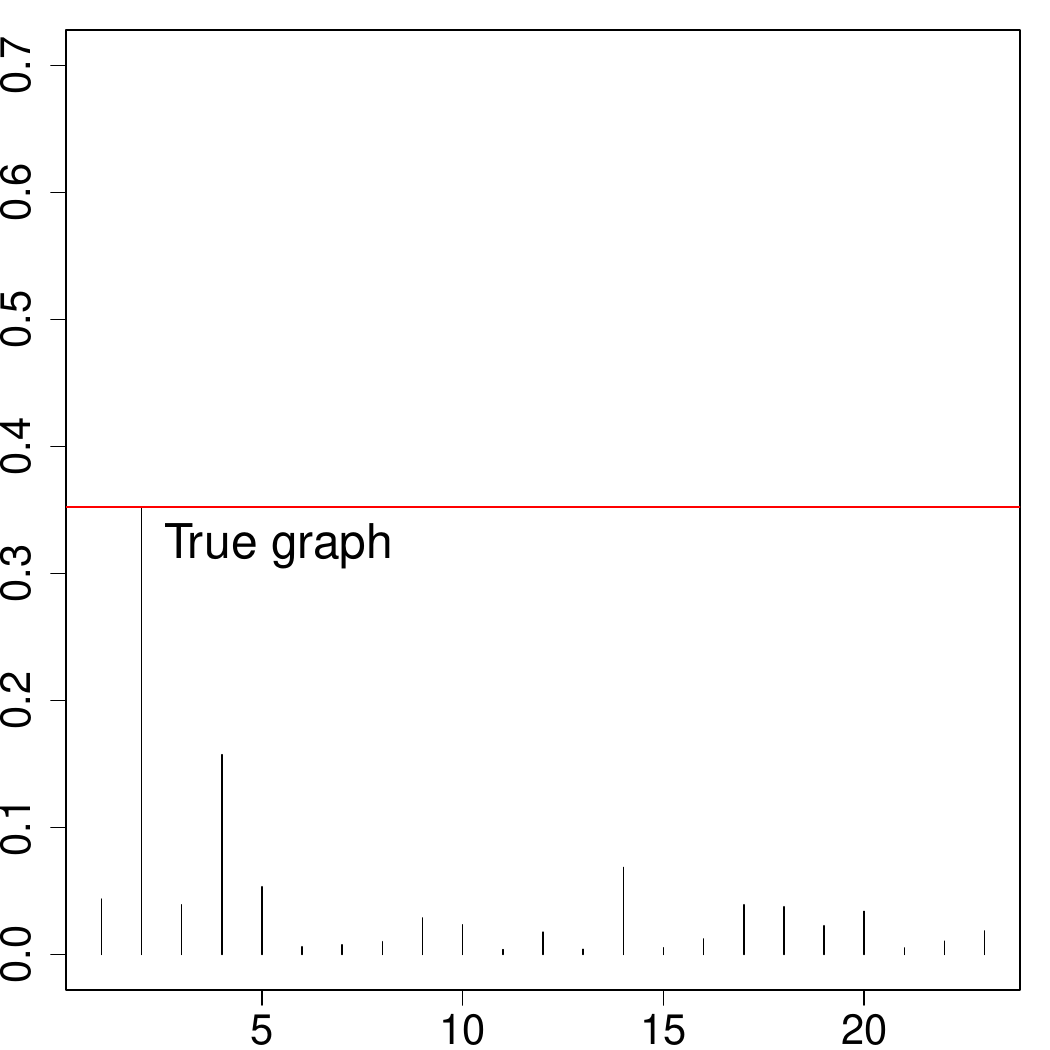}
    \label{fig:6node-post-subfig3}
}
\caption[]{Plot of the estimated posterior probability of graphs in simulation example \ref{example 1}; BDMCMC algorithm in \ref{fig:6node-post-subfig1}, Lenkoski algorithm \citep{lenkoski2013direct} in \ref{fig:6node-post-subfig2}, and WL algorithm \citep{wang2012efficient} in \ref{fig:6node-post-subfig3}.}
\label{fig:6node-post}
\end{figure}

Figure \ref{fig:6node-post} reports the estimated posterior distribution of the graphs for BDMCMC, WL, and Linkoski algorithm, respectively. Figure \ref{fig:6node-post-subfig1} indicates that our algorithm visited around $450$ different graphs and the estimated posterior distribution of the true graph is $0.66$, which is the graph with the highest posterior probability. Figure \ref{fig:6node-post-subfig2} shows that the Lenkoski algorithm visited around $400$ different graphs and the estimated posterior distribution of the true graph is $0.40$. Figure \ref{fig:6node-post-subfig3} shows that the WL algorithm visited only $23$ different graphs and the estimated posterior distribution of the true graph is $0.35$. 

To assess the performance of the graph structure, we compute the posterior probability of the true graph, and the calibration error (CE) measure, defined as follows
\begin{eqnarray}
\label{phat-error}
CE = \sum_{e \in \mathcal{W}}{|\hat{p}_e - I(e \in G_{true})|},
\end{eqnarray}
where, for each $e \in \mathcal{W}$, $\hat{p}_e$ is the posterior pairwise edge inclusion probability in (\ref{posterior-edge}) and $G_{true}$ is the true graph. The CE is positive with a minimum at $0$ and smaller is better.

Table \ref{table:6node} reports comparisons of our method with two other Bayesian approaches (WL and Lenkoski), reporting the mean values and standard errors in parentheses. We repeat the entire simulation $50$ times. The first and second columns show the performance of the algorithms. Our algorithm performs better due to its faster convergence feature. The third column shows the acceptance probability ($\alpha$) which is the probability of moving to a new graphical model. The fourth column shows that our algorithm is slower than the Lenkoski algorithm and faster than the WL approach. It can be argued that to make a fair comparison our method takes $60,000$ samples in 25 minutes while e.g. Lenkoski algorithm in 14 minutes takes only $60,000 \times 0.058 = 3,480$ efficient samples. For fair comparison we performed all simulations in R. However, our package {\tt BDgraph} efficiently implements the algorithm with C++ code linked to R. For $60,000$ iterations, our C++ code takes only 17 seconds instead of 25 minutes in R, which means around 90 times faster than the R code. It makes our algorithm computationally feasible for high-dimensional graphs.
 
\begin{table}[ht] \scriptsize 
\centering
\begin{tabular}{l*{8}{l}r}
\toprule[0.09 em]
             & P(true G $|$ data)  &    CE           & $\alpha$           & CPU time (min)     \\
\midrule
BDMCMC       & 0.66 (0.00)         & 0.47 (0.01)   & 1                    & 25 (0.14)           \\
Lenkoski     & 0.36 (0.02)         & 1.17 (0.08)   & 0.058 (0.001)        & 14 (0.13)         \\
WL           & 0.33 (0.12)         & 1.25 (0.46)   & 0.003 (0.0003)       & 37 (0.64)          \\
\toprule[0.09 em]
\end{tabular}
\caption{\label{table:6node} 
Summary of performance measures in simulation example \ref{example 1} for BDMCMC approach, Lenkoski \citep{lenkoski2013direct}, and WL \citep{wang2012efficient}. The table presents the average posterior probability of the true graph, the average calibration error (CE) which is defined in (\ref{phat-error}), the average acceptance probability ($\alpha$), and the average computing time in minutes, with $50$ replications and standard deviations in parentheses.}
\end{table}

\subsubsection{Extensive comparison with Bayesian methods}
\label{example 2}

We perform here a comprehensive simulation with respect to different graph structures to evaluate the performance of our Bayesian method and compare it with two recently proposed trans-dimensional MCMC algorithms; WL \citep{wang2012efficient} and Lenkoski \citep{lenkoski2013direct}. Corresponding to different sparsity patterns, we consider 7 different kinds of synthetic graphical models:
\begin{itemize}
\item[1.] \textit{Circle:} A graph with $k_{ii}=1$, $k_{i,i-1}=k_{i-1,i}=0.5$, and $k_{1p}=k_{p1}=0.4$, and $k_{ij}=0$ otherwise.
\item[2.] \textit{Star:} A graph in which every node is connected to the first node, with $k_{ii}=1$, $k_{1i}=k_{i1}=0.1$, and $k_{ij}=0$ otherwise.
\item[3.] \textit{AR(1):} A graph with $\sigma_{ij}=0.7^{|i-j|}$.
\item[4.] \textit{AR(2):} A graph with $k_{ii}=1$, $k_{i,i-1}=k_{i-1,i}=0.5$, and $k_{i,i-2}=k_{i-2,i}=0.25$, and $k_{ij}=0$ otherwise.
\item[5.] \textit{Random:} A graph in which the edge set $E$ is randomly generated from independent Bernoulli distributions with probability $2/(p-1)$ and the corresponding precision matrix is generated from $K \sim W_G(3,I_p)$.
\item[6.] \textit{Cluster:} A graph in which the number of clusters is $ \max \left\{ 2, \left[ p/20 \right] \right\} $. Each cluster has the same structure as a random graph. The corresponding precision matrix is generated from $K \sim W_G(3,I_p)$.
\item[7.] \textit{Scale-free:} A graph which is generated by using the B-A algorithm \citep{albert2002statistical}. The resulting graph has $p - 1$ edges. The corresponding precision matrix is generated from $K \sim W_G(3,I_p)$.
\end{itemize}

For each graphical model, we consider four different scenarios: (1) dimension $p = 10$ and sample size $n = 30$, (2) $p = 10$ and $n = 100$, (3) $p = 50$ and $n = 100$, (4) $p = 50$ and $n = 500$. 

For each generated sample, we fit our Bayesian method and two other Bayesian approaches (WL and Lenkoski) with a uniform prior for the graph and the G-Wishart prior $W_G(3, I_p)$ for the precision matrix. We run those three algorithms with the same starting points with $60,000$ iterations and $30,000$ as a burn in. Computation for this example was performed in parallel on $235$ batch nodes with $12$ cores and $24$ GB of memory, running Linux.

To assess the performance of the graph structure, we compute the calibration error (CE) measure defined in (\ref{phat-error}) and the $F_1$-score measure \citep{baldi2000assessing, powers2011evaluation} which is defined as follows
\begin{eqnarray}
\label{f1}
F_1\mbox{-score} = \frac{2 \mbox{TP}}{2 \mbox{TP + FP + FN}},
\end{eqnarray}
where TP, FP, and FN are the number of true positives, false positives, and false negatives, respectively. The $F_1$-score lies between $0$ and $1$, where $1$ stands for perfect identification and $0$ for bad identification.

Table \ref{table:F1-phat} reports comparisons of our method with two other Bayesian approaches, where we repeat the experiments 50 times and report the average $F_1$-score and CE with their standard errors in parentheses. Our method performs well overall as its $F_1$-score and its CE are the best in most of the cases, mainly because of its fast convergence rate. Both our method and the Lenkoski approach perform better compared to the WL approach. The main reason is that the WL approach uses a double Metropolis-Hastings (based on a block Gibbs sampler), which is an approximation of the exchange algorithm. On the other hand, both our method and the Lenkoski approach use the exchange algorithm based on exact sampling from the precision matrix. As we expected, the Lenkoski approach converges slower compared to our method. The reason seems to be the dependence of the Lenkoski approach on the choice of the tuning parameter, $\sigma_g^2$ \citep[step 3 in algorithm p. 124]{lenkoski2013direct}. In our simulation, we found that the convergence rate (as well acceptance probability) of the Lenkoski algorithm depends on the choice of $\sigma_g^2$. Here we choose  $\sigma_g^2 = 0.1$ as a default. From a theoretical point of view, both our BDMCMC and the Lenkoski algorithms converge to the true posterior distribution, if we run them a sufficient amount of time. Thus, the results from this table just indicate how quickly the algorithms converge.

\renewcommand{\tabcolsep}{5pt} 
\begin{table}[H]  \scriptsize 
\centering
\begin{tabular}{l*{8}{l}l}
\toprule
            \!& \multicolumn{3}{c}{$F_1$-score}                                & \multicolumn{4}{c}{CE}                            \\
            \cmidrule{2-4}                                                \cmidrule{6-8}      
            \!& \!BDMCMC            & Lenkoski          & WL               \!&\!&\!BDMCMC               \!&\!Lenkoski        \!&\!WL         \\
\midrule                                                                                                          
															                      \\
\multicolumn{2}{l}{p=10 \& n=30}                                                                                                           \\
															                      \\
circle     \!&\!\textbf{0.95} (0.00)& 0.93 (0.01)        & 0.24 (0.01)    \!&\!&\!\textbf{2.5} (1.6)    \!&\!4.9 (2.1)    \!&\!15.8 (5)     \\
star       \!& \!0.15 (0.02)        &\textbf{0.17} (0.02)& 0.16 (0.01)    \!&\!&\!\textbf{11.3} (2.1)   \!&\!14 (1.4)     \!&\!13.6 (3.4)     \\
AR1        \!&\!\textbf{0.90} (0.01)& 0.70 (0.01)        & 0.34 (0.02)    \!&\!&\!\textbf{4.4} (2.2)    \!&\!9.7 (2.4)    \!&\!12.5 (8.8)      \\
AR2        \!& \!0.56 (0.01)        &\textbf{0.59} (0.02)& 0.36 (0.01)    \!&\!&\!\textbf{11.5} (3.5)   \!&\!12.8 (3.1)   \!&\!16.3 (6.2)     \\
random     \!&\!\textbf{0.57} (0.03)& 0.50 (0.01)        & 0.34 (0.01)    \!&\!&\!\textbf{11.4} (8.0)   \!&\!15.3 (6.3)   \!&\!14.1 (14.0)     \\
cluster    \!&\!\textbf{0.61} (0.02)& 0.49 (0.01)        & 0.33 (0.01)    \!&\!&\!\textbf{10.3} (9.4)   \!&\!14.3 (7.3)   \!&\!13.5 (9.8)     \\
scale-free \!&\!\textbf{0.53} (0.03)& 0.45 (0.02)        & 0.31 (0.02)    \!&\!&\!\textbf{11.8} (8.8)   \!&\!15.6 (6.9)   \!&\!13.3 (6.5)     \\
                                                                                                                                           \\
\multicolumn{2}{l}{p=10 \& n=100}                                                                                                          \\
                                                                                                                                           \\
circle     \!&\!\textbf{0.99} (0.00)& 0.98 (0.00)        & 0.26 (0.01)        \!&\!&\!\textbf{1.0} (0.4)\!&\!2.2 (0.5)    \!&\!15.6 (6.7)         \\ 
star       \!&\!0.21 (0.02)         & 0.18 (0.02)        &\textbf{0.25} (0.02)\!&\!&\!\textbf{9.3} (1.6)\!&\!11.4 (1.3)   \!&\!11.4 (3.4)          \\
AR1        \!&\!\textbf{0.98} (0.00)& 0.95 (0.00)        & 0.34 (0.01)        \!&\!&\!\textbf{1.5} (0.4)\!&\!5.2 (0.5)    \!&\!13.0 (5.7)          \\ 
AR2        \!& \!0.89 (0.01)        &\textbf{0.90} (0.01)& 0.47 (0.01)        \!&\!&\!\textbf{4.1} (3.7)\!&\!5.6 (2.7)    \!&\!14.0 (7.5)         \\ 
random     \!&\!\textbf{0.76} (0.01)& 0.65 (0.02)        & 0.35 (0.01)        \!&\!&\!\textbf{7.0} (5.6)\!&\!10.7 (6.3)   \!&\!13.8 (10.5)         \\ 
cluster    \!&\!\textbf{0.74} (0.02)& 0.67 (0.02)        & 0.37 (0.02)        \!&\!&\!\textbf{6.4} (7.2)\!&\!9.9 (7.8)    \!&\!12.4 (9.4)        \\ 
scale-free \!&\!\textbf{0.69} (0.02)& 0.56 (0.02)        & 0.33 (0.02)        \!&\!&\!\textbf{7.9} (8.0)\!&\!11.6 (7.0)   \!&\!13.0 (7.8)          \\ 
                                                                                                                                           \\
\multicolumn{2}{l}{p=50 \& n=100}                                                                                                          \\
                                                                                                                                           \\
circle     \!&\!\textbf{0.99} (0.01)& 0.55 (0.10)     & 0.00 (0.00)    \!&\!&\!\textbf{2.5}(0.9)   \!&\!75.3 (7.2)    \!&\!50 (0.0)     \\
star       \!&\!\textbf{0.17} (0.04)& 0.09 (0.04)     & 0.00 (0.00)    \!&\!&\!68.8 (4.4)        \!&\!166.7 (4.5)    \!&\!\textbf{49} (0.0)     \\
AR1        \!&\!\textbf{0.86} (0.04)& 0.33 (0.09)     & 0.00 (0.00)    \!&\!&\!\textbf{19.0} (4.5)\!&\!159.1 (5.4)   \!&\!49 (0.0)     \\
AR2        \!&\!\textbf{0.86} (0.04)& 0.49 (0.17)     & 0.00 (0.00)    \!&\!&\!\textbf{28.6} (5.7) \!&\!117.5 (5.4)   \!&\!97 (0.0)     \\
random     \!&\!\textbf{0.51} (0.09)& 0.21 (0.07)     & 0.00 (0.00)    \!&\!&\!73.2 (18.5)        \!&\!250.6 (36.7)    \!&\!\textbf{49.2} (5.6)     \\
cluster    \!&\!\textbf{0.55} (0.11)& 0.18 (0.07)     & 0.00 (0.00)    \!&\!&\!72.8 (18.2)        \!&\!246.0 (44.4)    \!&\!\textbf{47.8} (8.4)     \\
scale-free \!&\!\textbf{0.49} (0.11)& 0.19 (0.07)     & 0.00 (0.00)    \!&\!&\!72.4 (22.5)        \!&\!243.1 (47.8)    \!&\!\textbf{49} (0.0)     \\
                                                                                                                           \\
\multicolumn{2}{l}{p=50 \& n=500}                                                                                          \\
                                                                                                                           \\
circle     \!&\!\textbf{1.00} (0.01)& 0.72 (0.09)     & 0.00 (0.00)    \!&\!&\!\textbf{1.7} (0.6)   \!&\!55.8 (5.4)   \!&\!50 (0.0)     \\
star       \!&\!\textbf{0.65} (0.05)& 0.35 (0.05)     & 0.00 (0.00)    \!&\!&\!\textbf{31.7} (4.5)  \!&\!92.3 (3.6)   \!&\!49 (0.0)     \\
AR1        \!&\!\textbf{0.94} (0.02)& 0.54 (0.07)     & 0.00 (0.00)    \!&\!&\!\textbf{7.2} (1.9)   \!&\!84.9 (4.0)   \!&\!49 (0.0)     \\
AR2        \!&\!\textbf{0.98} (0.01)& 0.78 (0.11)     & 0.00 (0.00)    \!&\!&\!\textbf{4.8} (1.8)   \!&\!61.7 (4.4)   \!&\!97 (0.0)     \\
random     \!&\!\textbf{0.73} (0.09)& 0.34 (0.10)     & 0.00 (0.00)    \!&\!&\!\textbf{34.3} (11.2) \!&\!149.3 (28.8)   \!&\!50.7 (7.0)     \\
cluster    \!&\!\textbf{0.74} (0.09)& 0.32 (0.13)     & 0.00 (0.00)    \!&\!&\!\textbf{32.2} (10.6) \!&\!142.2 (27.3)   \!&\!48.5 (5.9)     \\
cale-free \!&\!\textbf{0.73} (0.10)& 0.33 (0.08)     & 0.00 (0.00)    \!&\!&\!\textbf{35.3} (13.6) \!&\!151.7 (26.9)   \!&\!49 (0.0)     \\
\bottomrule
\end{tabular}
\caption{\label{table:F1-phat} 
Summary of performance measures in simulation example \ref{example 2} for BDMCMC approach, Lenkoski \citep{lenkoski2013direct}, and WL \citep{wang2012efficient}. The table presents the $F_1$-score, which is defined in (\ref{f1}) and CE, which is defined in (\ref{phat-error}), for different models with $50$ replications and standard deviations in parentheses. The $F_1$-score reaches its best score at 1 and its worst at 0. The CE is positive valued for which $0$ is minimum and smaller is better. The best models for both $F_1$-score and CE are boldfaced.}
\end{table}

Table \ref{table: time-alpha} reports the average running time and acceptance probability ($\alpha$) with their standard errors in parentheses across all 7 graphs with their 50 replications. It shows that our method compared to the Lenkoski approach is slower. The reason is that our method scans through all possible edges for calculating the birth/death rates, which is computationally expensive. On the other hand, in the Lenkoski algorithm, a new graph is selected by randomly choosing one edge which is computationally fast but not efficient. The table shows that the acceptance probability ($\alpha$) for both WL and Lenkoski is small especially for the WL approach. Note the $\alpha$ here is the probability that the algorithm moves to a new graphical model and it is not related to the double Metropolis-Hastings algorithm. The $\alpha$ in the WL approach is extremely small and it should be the cause of the approximation which has been used for the ratio of prior normalizing constants. As \citet{murray2012mcmc} pointed out these kinds of algorithms can suffer high rejection rates. For the Lenkoski approach the $\alpha$ is relatively small, but compared with the WL method is much better. As in the Lenkoski approach, a new graph is proposed by randomly choosing one edge, yielding a relatively small acceptance probability.   

\begin{table}[H]  \scriptsize 
\centering
\begin{tabular*}{\linewidth}{ @{\extracolsep{\fill}} ll *{13}c @{}}
\toprule
                          &            & BDMCMC             & Lenkoski             & WL                     \\
\midrule \addlinespace    
\multirow{2}{*}{p = 10}   & $\alpha$   & 1                  & 0.114 (0.001)        & 8.8e-06 (4.6e-11)      \\  
			  & Time       & 97 (628)           & 40 (225)             & 380 (11361)            \\
\midrule \addlinespace    
\multirow{2}{*}{p = 50}   & $\alpha$   &  1                 & 0.089 (0.045)        & 0.0000 (0.0000)        \\  
                          & Time       & 5408 (1694)        & 1193 (1000)          & 9650 (1925)            \\
\bottomrule
\end{tabular*}
\caption{\label{table: time-alpha} Comparison of our BDMCMC algorithm with the WL approach \citep{wang2012efficient} and Lenkoski approach \citep{lenkoski2013direct}. It presents the average computing time in minutes and the average probability of acceptance ($\alpha$) with their standard deviations in parentheses.}
\end{table}

\subsubsection{Comparison with frequentist methods}
\label{example 3}

We also compare the performance of our Bayesian method with two popular frequentist methods, the graphical lasso (glasso) \citep{friedman2008sparse} and Meinshausen-Buhlmann graph estimation (mb) \citep{meinshausen2006high}. We consider the same $7$ graphical models with the same scenarios in the previous example.

For each generated sample, we fit our Bayesian method with a uniform prior for the graph and the G-Wishart prior $W_G(3, I_p)$ for the precision matrix. To fit the glasso and mb methods, however, we must specify a regularization parameter $\lambda$ that controls the sparsity of the graph. The choice of $\lambda$ is critical since different $\lambda$'s may lead to different graphs. We consider the glasso method with three different regularization parameter selection approaches, which are the stability approach to regularization selection (stars) \citep{liu2010stability}, rotation information criterion (ric) \citep{zhao2012huge}, and the extended Bayesian information criterion (ebic) \citep{foygel2010ebic}. Similarly, we consider the mb method with two regularization parameter selection approaches, namely stars and the ric. We repeat all the experiments 50 times.

Table \ref{table F1-score} provides comparisons of all approaches, where we report the averaged $F_1$-score with their standard errors in parentheses. Our Bayesian approach performs well as its $F_1$-score typically out performs all frequentist methods, except in the unlikely scenario of a high number of observations where it roughly equals the performance of the mb method with stars criterion. All the other approaches appear to perform well in some cases, and fail in other cases. For instance, when $p = 50$, the mb method with ric is the best for the AR(1) graph and the worst for the circle graph. 

\renewcommand{\tabcolsep}{4pt} 
\begin{table}[H]  	\scriptsize 
\centering
\begin{tabular}{l*{8}{l}l}
\toprule
          &                       & \multicolumn{4}{c}{glasso}                             & \multicolumn{3}{c}{mb}                            \\
                                  \cmidrule{4-6}                                          \cmidrule{8-9}      
          & BDMCMC                && stars             & ric               & ebic             && stars             & ric         \\
\midrule                                                                                                          
															                      \\
\multicolumn{2}{l}{p=10 \& n=30}                                                                                                           \\
															                      \\
circle     &\textbf{0.95} (0.00)&& 0.00 (0.00)     & 0.01 (0.01)        &\textbf{0.48} (0.00)&& 0.42 (0.01)     & 0.01 (0.01)            \\
star       &\textbf{0.15} (0.02)&& 0.01 (0.00)     &\textbf{0.15} (0.02)& 0.00 (0.00)        && 0.01 (0.02)     & 0.14 (0.02)   \\
AR1        &\textbf{0.90} (0.01)&& 0.20 (0.13)     & 0.61 (0.01)        & 0.17 (0.07)        && 0.46 (0.01)     &\textbf{0.83} (0.01)           \\
AR2        &\textbf{0.56} (0.01)&& 0.09 (0.02)     &\textbf{0.19} (0.02)& 0.00 (0.00)        && 0.07 (0.02)     & 0.19 (0.02)            \\
random     &\textbf{0.57} (0.03)&& 0.36 (0.06)     & 0.48 (0.02)        & 0.08 (0.04)        && 0.45 (0.03)     &\textbf{0.53} (0.03)           \\
cluster    &\textbf{0.61} (0.02)&& 0.45 (0.05)     & 0.54 (0.02)        & 0.07 (0.04)        && 0.50 (0.02)     &\textbf{0.54} (0.02)            \\
scale-free &\textbf{0.53} (0.03)&& 0.30 (0.05)     & 0.4 (0.02)         & 0.06 (0.02)        && 0.36 (0.03)      &\textbf{0.46} (0.03)            \\
                                                                                                                                            \\
\multicolumn{2}{l}{p=10 \& n=100}                                                                                                           \\
                                                                                                                                            \\
circle     &\textbf{0.99} (0.00)&& 0.00 (0.00)     & 0.50 (0.08)        & 0.45 (0.00)     &&\textbf{0.89} (0.08)& 0.81 (0.09)           \\ 
star       & 0.21 (0.02)        && 0.08 (0.02)     &\textbf{0.29} (0.03)& 0.01 (0.00)     && 0.07 (0.03)        &\textbf{0.29} (0.03)   \\ 
AR1        &\textbf{0.98} (0.00)&& 0.90 (0.01)     & 0.57 (0.00)        & 0.56 (0.00)     &&\textbf{0.94} (0.00)& 0.85 (0.00)           \\ 
AR2        &\textbf{0.89} (0.01)&& 0.34 (0.06)     & 0.63 (0.00)        & 0.08 (0.05)     && 0.41 (0.01)        &\textbf{0.64} (0.01)           \\ 
random     &\textbf{0.76} (0.01)&& 0.61 (0.02)     & 0.57 (0.01)        & 0.45 (0.07)     &&\textbf{0.68} (0.02)& 0.61 (0.02)           \\ 
cluster    &\textbf{0.74} (0.02)&& 0.66 (0.03)     & 0.59 (0.02)        & 0.53 (0.07)     &&\textbf{0.68} (0.03)& 0.61 (0.03)            \\ 
scale-free &\textbf{0.69} (0.02)&& 0.56 (0.02)     & 0.48 (0.008)       & 0.34 (0.07)     &&\textbf{0.63} (0.02)& 0.52 (0.02)           \\ 
                                                                                                                                               \\
\multicolumn{2}{l}{p=50 \& n=100}                                                                                                              \\
                                                                                                                                               \\
circle     &\textbf{0.99} (0.01)&& 0.28 (0.05)        & 0.00 (0.00)     &\textbf{0.28} (0.01)&& 0.00 (0.00)        & 0.00 (0.00)            \\
star       &\textbf{0.17} (0.04)&& 0.14 (0.06)        & 0.06 (0.05)     & 0.00 (0.00)        &&\textbf{0.15} (0.04)& 0.05 (0.041)            \\
AR1        &\textbf{0.86} (0.04)&& 0.56 (0.04)        & 0.59 (0.03)     & 0.49 (0.05)        && 0.82 (0.02)        &\textbf{0.98} (0.02)    \\
AR2        &\textbf{0.86} (0.04)&& 0.59 (0.02)        & 0.02 (0.02)     & 0.00 (0.00)        &&\textbf{0.66} (0.02)& 0.02 (0.02)            \\
random     & 0.51 (0.09)        &&\textbf{0.52} (0.10)& 0.40 (0.16)     & 0.04 (0.13)        &&\textbf{0.61} (0.21)& 0.49 (0.21)             \\
cluster    &\textbf{0.55} (0.11)&& 0.54 (0.06)        & 0.42 (0.18)     & 0.13 (0.24)        &&\textbf{0.64} (0.22)& 0.50 (0.22)            \\
scale-free &\textbf{0.49} (0.11)&& 0.48 (0.10)        & 0.32 (0.18)     & 0.02 (0.09)        &&\textbf{0.60} (0.23)& 0.40 (0.23)            \\
                                                                                                                                                \\
\multicolumn{2}{l}{p=50 \& n=500}                                                                                                               \\
                                                                                                                                                \\
circle     &\textbf{1.00} (0.01)&&\textbf{0.27} (0.05)& 0.00 (0.00)     & 0.25 (0.01)     && 0.00 (0.00)        & 0.00 (0.00)                \\
star       &\textbf{0.65} (0.05)&& 0.29 (0.12)        & 0.60 (0.07)     & 0.01 (0.02)     && 0.31 (0.07)        &\textbf{0.60} (0.07)                \\
AR1        & 0.94 (0.02)        && 0.57 (0.02)        & 0.54 (0.02)     & 0.44 (0.02)     &&\textbf{0.97} (0.02)&\textbf{0.98} (0.01)         \\
AR2        &\textbf{0.98} (0.01)&& 0.69 (0.03)        & 0.64 (0.01)     & 0.66 (0.04)     &&\textbf{0.89} (0.02)& 0.69 (0.02)                \\
random     &\textbf{0.73} (0.09)&& 0.62 (0.12)        & 0.46 (0.15)     & 0.56 (0.13)     &&\textbf{0.82} (0.24)& 0.61 (0.24)             \\
cluster    &\textbf{0.74} (0.09)&& 0.65 (0.10)        & 0.51 (0.17)     & 0.58 (0.10)     &&\textbf{0.82} (0.23)& 0.64 (0.23)             \\
scale-free &\textbf{0.73} (0.10)&& 0.57 (0.14)        & 0.41 (0.15)     & 0.47 (0.15)     &&\textbf{0.82} (0.24)& 0.62 (0.24)             \\
\bottomrule
\end{tabular}
\caption{
Summary of performance measures in simulation example \ref{example 3} for BDMCMC approach, glasso \citep{friedman2008sparse} with 3 criteria and mb \citep{meinshausen2006high} method with 2 criteria. The table reports \label{table F1-score} $F_1$-score, which is defined in (\ref{f1}), for different models with $50$ replications and standard deviations are in parentheses. The $F_1$-score reaches its best score at 1 and its worst at 0. The two top models are boldfaced.}
\end{table}

To assess the performance of the precision matrix estimation, we use the Kullback-Leibler divergence \citep{kullback1951information} which is given as follows
\begin{eqnarray}
\label{kl}
\mbox{KL} = \frac{1}{2} \left[ \mbox{tr} \left( K^{-1}_{\mbox{true}} \hat{K} \right) - p - \log \left( \frac{|\hat{K}|}{|K_{\mbox{true}}|} \right) \right],
\end{eqnarray}
where $K_{\mbox{true}}$ is the true precision matrix and $\hat{K}$ is the estimate of the precision matrix. 

Table \ref{table KL} provides a comparison of all methods, where we report the averaged KL with their standard errors in parentheses. Based on KL, the overall performance of our Bayesian approach is good as its KL is the best in all scenarios except one.  

\renewcommand{\tabcolsep}{6pt} 
\begin{table}[H]  	\scriptsize 
\centering
\begin{tabular}{l*{8}{l}l}
\toprule
          &                        &   \multicolumn{4}{c}{glasso}                                           \\
                                   \cmidrule{4-6}                    
          & BDMCMC                && stars             & ric               & ebic      \\
\midrule                                                                             
														 \\
\multicolumn{2}{l}{p=10 \& n=30}                                                                              \\
														 \\
circle     &\textbf{0.73} (0.12)  && 15.84 (0.03)         & - -                  & 10.34 (1.33)           \\
star       & 0.57 (0.08)          && 0.31 (0.00)          &\textbf{0.22} (0.00)  & 0.33 (0.01)          \\
AR1        &\textbf{0.70} (0.10)  && 3.63 (0.07)          & 1.59 (0.06)          & 2.77 (2.34)          \\
AR2        &\textbf{1.22} (0.07)  && 1.27 (0.00)          & 1.26 (0.00)          & 1.28 (0.00)          \\
random     &\textbf{0.67} (0.08)  && 8.32 (305)           & - -                  & 12.44 (1637)           \\
cluster    &\textbf{0.61} (0.06)  && 4.90 (2.37)          & 3.74 (3.23)          & 5.72 (7.35)           \\
scale-free &\textbf{0.65} (0.07)  && 5.83 (12.35)         & - -                  & 6.59 (26.62)          \\
                                                                                                               \\
\multicolumn{2}{l}{p=10 \& n=100}                                                                              \\
                                                                                                               \\
circle     & \textbf{0.14} (0.00) && 15.95 (0.01)         & - -                  & 9.60 (0.56)          \\
star       & 0.13 (0.00)          && 0.15 (0.00)          & \textbf{0.10} (0.00) & 0.17 (0.00)          \\
AR1        & \textbf{0.12} (0.00) && 2.88 (0.16)          & 0.81 (0.01)          & 0.37 (0.00)          \\
AR2        & \textbf{0.28} (0.01) && 1.24 (0.01)          & 1.14 (0.00)          & 1.25 (0.02)          \\
random     & \textbf{0.16} (0.00) && 4.47 (1.09)          & 3.30 (0.76)          & 3.92 (2.55)          \\
cluster    & \textbf{0.13} (0.00) && 4.46 (12.62)         & 3.62 (8.17)          & 4.47 (30.31)          \\
scale-free & \textbf{0.16} (0.00) && 4.14 (1.27)          & 3.01 (0.70)          & 3.68 (1.94)          \\
                                                                                                               \\
\multicolumn{1}{c}{p=50 \& n=100}                                                                              \\
                                                                                                               \\
circle     &\textbf{0.67} (0.13)  && 117.12 (32.12)      & - -            & 115.88 (5.88)          \\
star       & 1.75 (0.21)          &&\textbf{1.05} (0.08) & 1.27 (0.07)    & 1.49 (0.16)          \\
AR1        &\textbf{1.17} (0.23)  && 8 (1.10)            & 8.92 (0.50)    & 6.20 (0.93)          \\
AR2        &\textbf{1.97} (0.33)  && 6.56 (0.19)         & 7.29 (0.06)    & 7.27 (0.09)          \\
random     &\textbf{2.01} (0.42)  && 20.83 (6.44)        & - -            & 30.07 (12.54)          \\
cluster    &\textbf{1.94} (0.42)  && 19.82 (3.75)        & - -            & 26.47 (5.08)             \\
scale-free &\textbf{1.96} (0.45)  && 20.91 (6.71)        & - -            & 28.98 (8.40)             \\
                                                                                                             \\
\multicolumn{1}{c}{p=50 \& n=500}                                                                            \\
                                                                                                             \\
circle     &\textbf{0.11} (0.01)  && 111.75 (30.56)     & - -             & 111.32 (3.26)      \\
star       &\textbf{0.34} (0.04)  && 0.78 (0.04)        & 0.63 (0.03)     & 0.96 (0.06)        \\
AR1        &\textbf{0.15} (0.02)  && 4.87 (0.45)        & 3.51 (0.23)     & 1.75 (0.08)        \\
AR2        &\textbf{0.19} (0.03)  && 5.50 (0.2)         & 6.42 (0.13)     & 4.01 (0.10)        \\
random     &\textbf{0.26} (0.06)  && 18.89 (5.58)       & 17.14 (6.59)    & 17.80 (8.79)       \\
cluster    &\textbf{0.24} (0.05)  && 21.09 (21.51)      & - -             & 21.34 (28.62)       \\
scale-free &\textbf{0.25} (0.07)  && 19.86 (7.69)       & - -             & 19.61 (16.65)       \\
\bottomrule
\end{tabular}
\caption{\label{table KL} 
Summary of performance measures in simulation example \ref{example 3} for BDMCMC approach and glasso \citep{friedman2008sparse} with 3 criteria. The table reports the KL measure, which is defined in (\ref{kl}), for different models with $50$ replications and standard deviations are in parentheses. The KL is positive valued for which $0$ is minimum and smaller is better. The best models are boldfaced.}
\end{table}

\subsection{Application to human gene expression data} 
\label{example: human gene data}

We apply our proposed method to analyze the large-scale human gene expression data which was originally described by \citet{bhadra2013joint, chen2008considering}, and \citet{stranger2007population}. The data are collected by \citet{stranger2007population} using Illumina's Sentrix Human-6 Expression BeadChips to measure gene expression in B-lymphocyte cells from Utah (CEU) individuals of Northern and Western European ancestry. They consider 60 unrelated individuals whose genotypes are available from the Sanger Institute website (\url{ftp://ftp.sanger.ac.uk/pub/genevar}). The genotype is coded as 0, 1, and 2 for rare homozygous, heterozygous and homozygous common alleles. Here the focus is on the $3125$ Single Nucleotide Polymorphisms (SNPs) that have been found in the 5' UTR (untranslated region) of mRNA (messenger RNA) with a minor allele frequency $\geq 0.1$. There were four replicates for each individual. Since the UTR has been subject to investigation previously, it should have an important role in the regulation of the gene expression. The raw data were background corrected and then quantile normalized across replicates of a single individual and then median normalized across all individuals. We chose the 100 most variable probes among the 47,293 total available probes corresponding to different Illumina TargetID. Each selected probe corresponds to a different transcript. Thus, we have n = 60 and p = 100. The data are available in the R package, {\tt BDgraph}. \citet{bhadra2013joint} have analyzed the data by adjusting the effect of SNPs using an expression quantitative trait loci (eQTL) mapping study.  They found 54 significant interactions among the 100 traits considered. Previous studies have shown that these data are an interesting case study to carry out prediction.

We place a uniform distribution as an uninformative prior on the graph and the G-Wishart $W_G(3,I_{100})$ on the precision matrix. We run our BDMCMC algorithm for $60,000$ iterations with a $30,000$ sweeps burn-in. 

\begin{figure}[!ht]
\vspace{-4em} %
\centering
\includegraphics[width=0.8\textwidth]{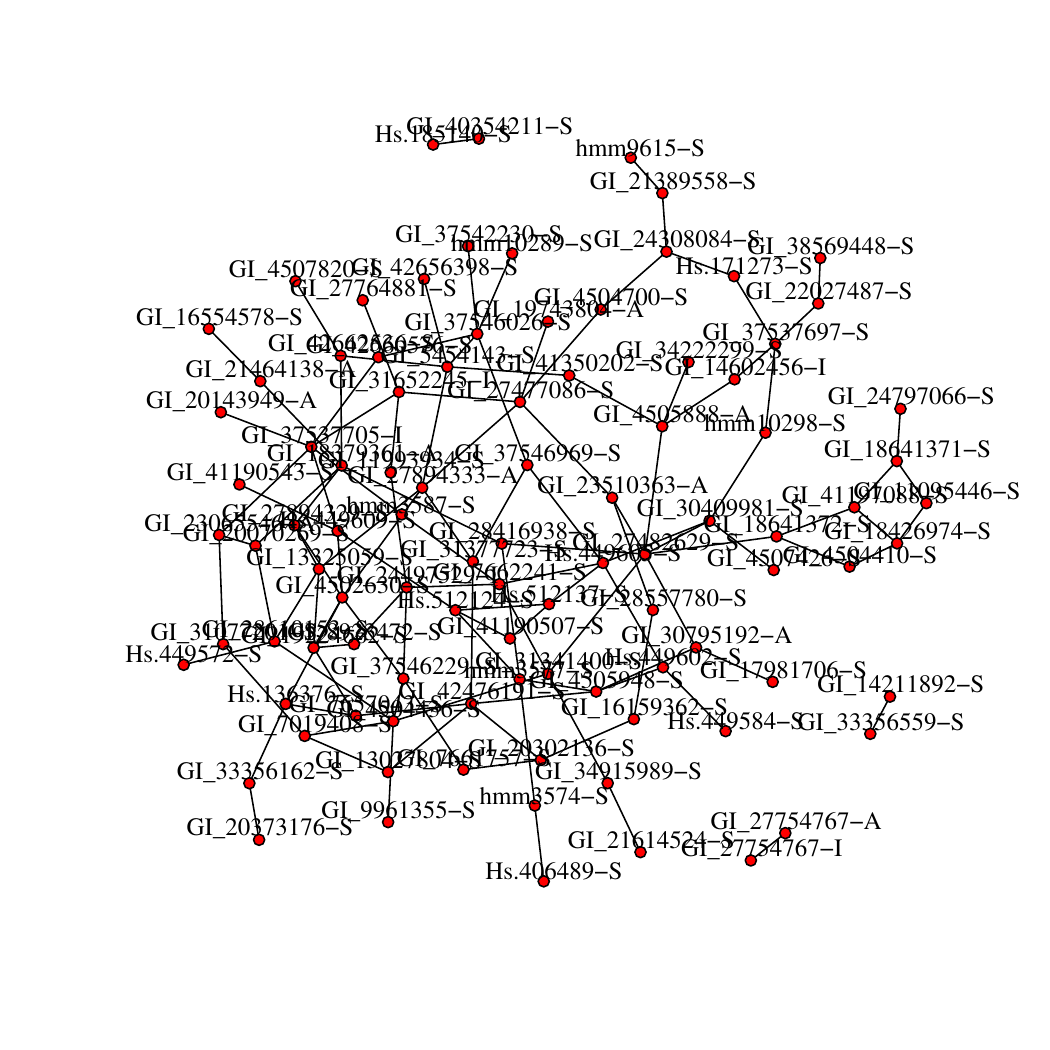}
\caption{ \label{fig:graph-gene data} The inferred graph for the human gene expression  data set. It reports the selected graph with $86$ significant edges for which their posterior inclusion probabilities (\ref{posterior-edge}) are more then $0.6$.}
\end{figure}

\begin{figure}[!ht]
\vspace{-2em} %
\centering
\includegraphics[width=0.5\textwidth]{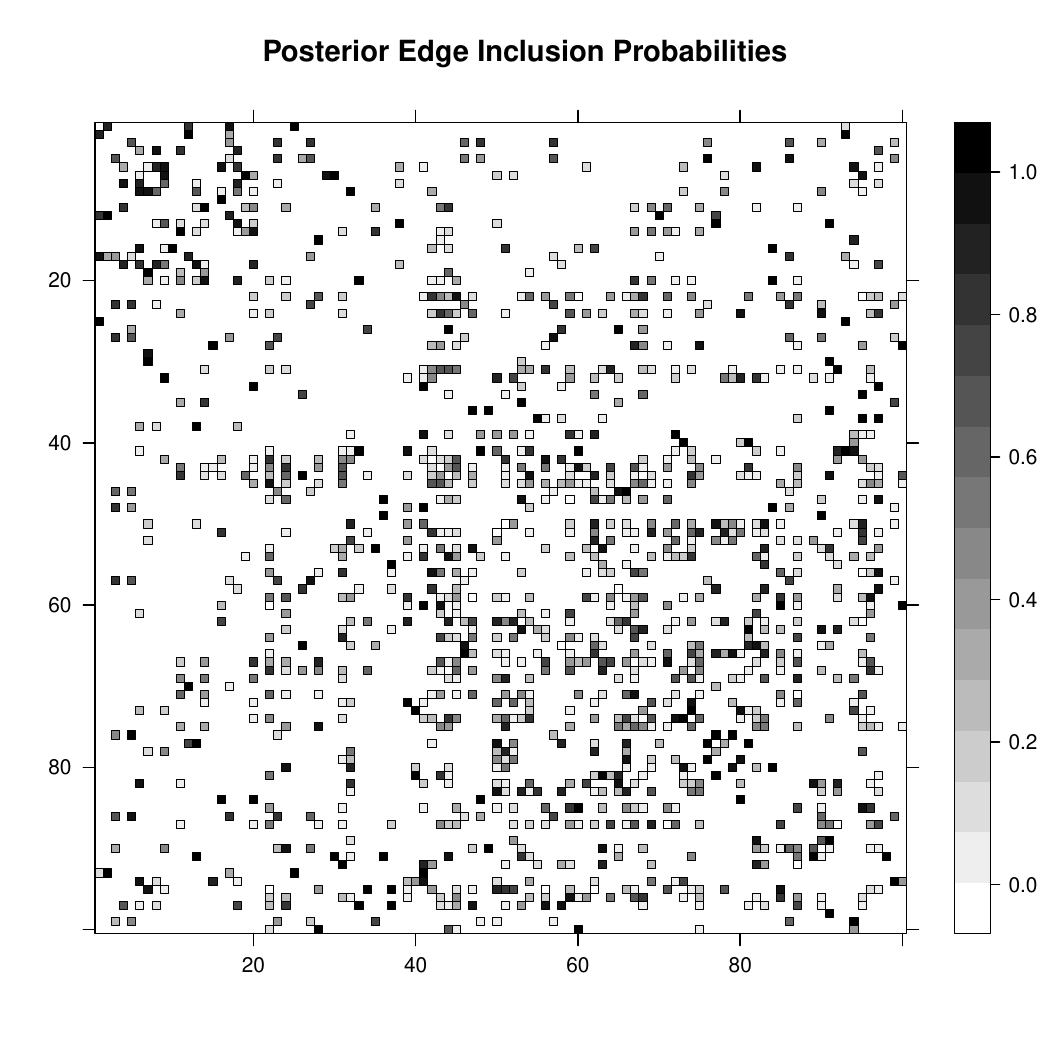}
\caption{ \label{fig:phat-gene data} Image visualization of the posterior pairwise edge inclusion probabilities for all possible edges in the graph.}
\end{figure}

The graph with the highest posterior probability is the graph with $281$ edges, which includes almost all the significant interactions discovered by \citet{bhadra2013joint}. Figure \ref{fig:graph-gene data} shows the selected graph with $86$ edges, for which the posterior inclusion probabilities in (\ref{posterior-edge}) is greater then $0.6$. Edges in the graph show the interactions among the genes. Figure \ref{fig:phat-gene data} shows the image of the the all posterior inclusion probabilities for visualization.

\subsection{Extension to time course data} 
\label{example: mammary data}

Here, to demonstrate how well our proposed methodology can be extended to other types of graphical models, we focus on graphical models involving time series data \citep{dahlhaus2003causality, abegaz2013sparse}. We show how graphs can be useful in modeling real-world problems such as gene expression time course data.

Suppose we have a $T$ time point longitudinal microarray study across $p$ genes. We assume a stable dynamic graph structure for the time course data as follows:
\begin{eqnarray}
x_t \sim N_p(f(t), K^{-1}), \qquad \mbox{for} \ \ t = 1, ..., T,
\end{eqnarray}
in which vector $f(t)=\{ f_i(t) \}_{p}$ with $f_i(t) = \beta_i ' h(t) = \sum_{r=1}^{m} \beta_{ir} h_r(t)$, $\beta_i = (\beta_{i1}, ..., \beta_{im})'$, $h(t) = (h_1(t), ..., h_m(t))'$, and $m$ is the number of basic elements. $h(t)$ is a cubic spline basis which should be continuous with continuous first and second derivatives \citep[chapter 5]{hastie2009elements}. The aim of this model is to find a parsimonious description of both time dynamics and gene interactions.

For this model, we place a uniform distribution as an uninformative prior on the graph and a G-Wishart $W_G(3,I_{p})$ on the precision matrix. For a prior distribution for $\beta_i$, we choose $N_p(\mu_{0i}, B_{0i})$, with $i = 1,...,p$. Thus, based on our likelihood and priors, the conditional distribution of $\beta_i$ is 
\begin{eqnarray}
\beta_i | \xb, K, G \sim N_p(\mu_i, B_i),
\end{eqnarray}
in which
\begin{align*}
B_i &= ( B_{0i}^{-1}+K_{ii} \sum_{t=1}^{T} h(t) h^{T}(t)  )^{-1}, \\
\mu_i &= B_i ( B_{0i}^{-1} \mu_{0i} + \sum_{t=1}^{T} h(t) ( x_t ' K_{V,i} - K_{i,V \setminus i} f_{-i}(t) ) ).
\end{align*}
Thus, to account for the time effect in our model, we require one more Gibbs sampling step in the BDMCMC algorithm for updating $\beta_i$, for $i = 1,...,p$.

To evaluate the efficiency of the method, we focus on the mammary gland gene expression time course data from \citet{stein2004involution}. The data reports a large time course Affymetrix microarray experiment across different developmental stages performed by using mammary tissue from female mice. There are $12,488$ probe sets representing $\sim8600$ genes. In total, the probe sets are measured across $54$ arrays with three mice used at each of $18$ time points. The time points are in the four main stages, as follows: {\tt virgin}, 6, 10, and 12 weeks; {\tt pregnancy}, 1, 2, 3, 8.5, 12.5, 14.5, and 17.5; {\tt lactation}, 1, 3, and 7; {\tt involution}, 1, 2, 3, 4, and 20. By using cluster analysis, we identify $30$ genes which provide the best partition among the developmental stages. Those genes play a role in the transitions across the main developmental events. The mammary data is available in the R package {\tt smida}; for more details about the data see \citet[chapter one]{wit2004statistics}. \citet{abegaz2013sparse} analyze this data based on a sparse time series chain graphical model. By using our proposed methodology, we infer the interactions between the crucial genes.

\begin{figure}[!ht]
\vspace{-4em} %
\centering
\includegraphics[width=0.7\textwidth]{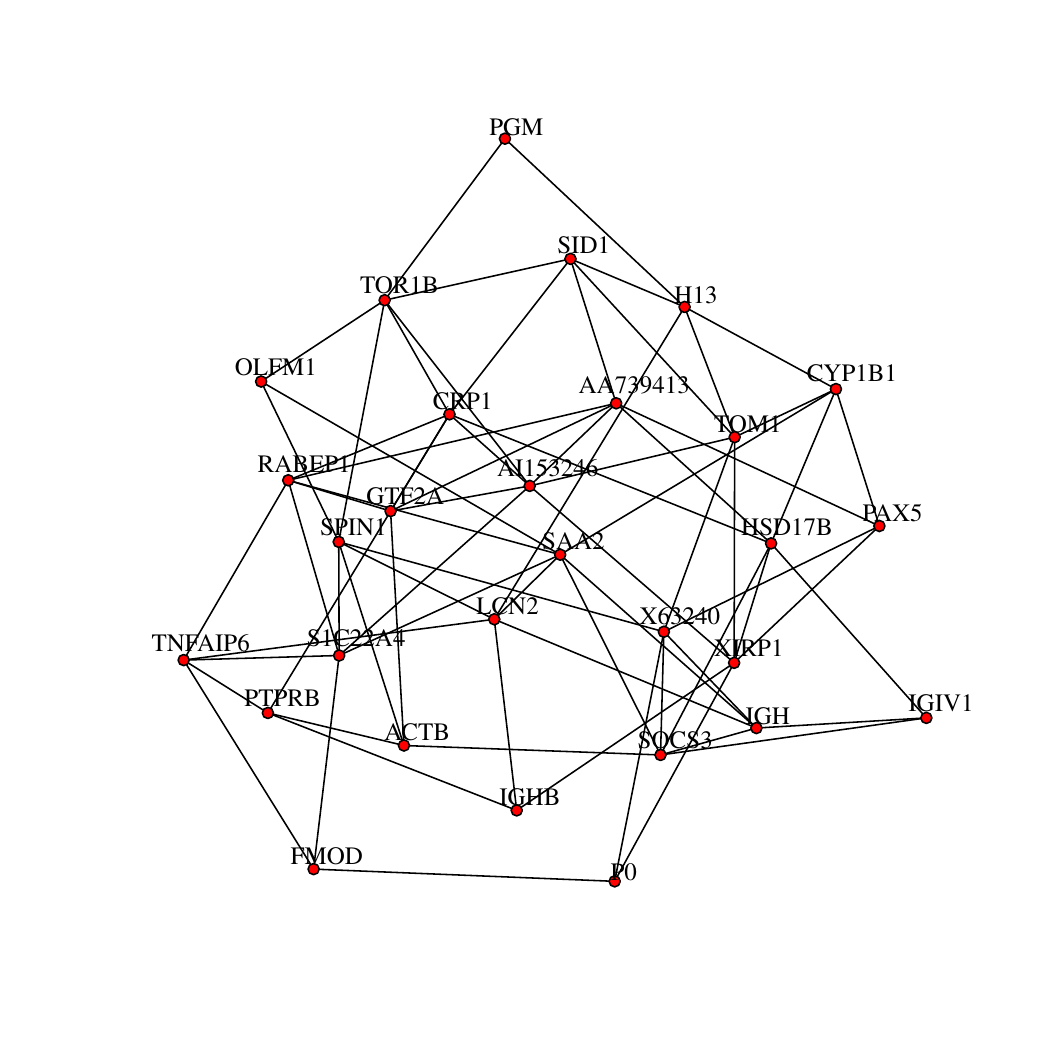}
\caption{ \label{fig:graph-mammary data} The inferred graph for the mammary gland gene expression data set. It reports the selected graph with $56$ significant edges for which their posterior inclusion probabilities (\ref{posterior-edge}) are more then $0.6$.}
\end{figure}

By placing a uniform prior on the graph and the G-Wishart $W_G(3,I_{30})$ on the precision matrix, we run our BDMCMC algorithm for $60,000$ iterations using $30,000$ as burn-in. Figure \ref{fig:graph-mammary data} shows the selected graph based on the output of our BDMCMC algorithm. The graph shows the edges with a posterior inclusion probability greater then $0.6$. As we can see in Figure \ref{fig:graph-mammary data}, the genes with the highest number of edges are LCN2, HSD17B, CRP1, and RABEP1, which each have $7$ edges. \citet{schmidt2007dual} suggested that gene LCN2 (lipocalin 2) plays an important role in the innate immune response to bacterial infection and also functions as a growth factor. For the gene HSD17B (17-$\beta$ hydroxysteroid dehydrogenase), past studies suggest that this gene family provides each cell with the necessary mechanisms to control the level of intracellular androgens and/or estrogens \citep{labrie1997key}. Gene CRP1 was identified by \citet{abegaz2013sparse} as a likely hub.

\begin{figure}[!ht]
\vspace{-1em} %
\centering
\includegraphics[width=0.6\textwidth]{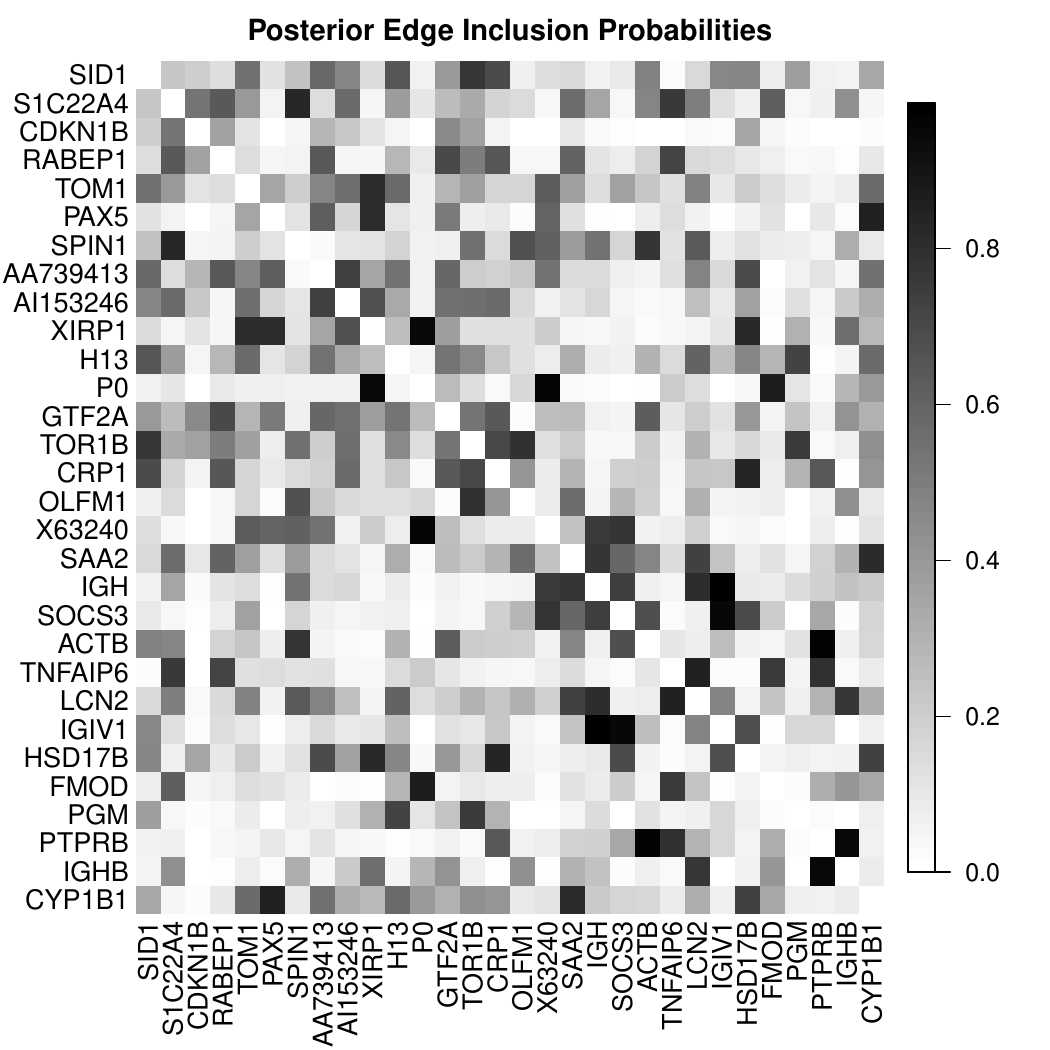}
\caption{ \label{fig:phat-mammary data} Image visualization of the posterior pairwise edge inclusion probabilities of all possible edges in the graph.}
\end{figure}

\section{Discussion}

We introduce a Bayesian approach for graph structure learning based on Gaussian graphical models using a trans-dimensional MCMC methodology. The proposed methodology is based on the birth-death process. In Theorem \ref{BD theorem}, we derived the conditions for which the balance conditions of the birth-death MCMC method holds. According to those conditions we proposed a convenient BDMCMC algorithm, whose stationary distribution is our joint posterior distribution. We showed that a scalable Bayesian method exists, which, also in the case of large graphs, is able to distinguish important edges from irrelevant ones and detect the true model with high accuracy. The resulting graphical model is reasonably robust to modeling assumptions and the priors used.

As we have shown in our simulation studies (\ref{example 1}), in Gaussian graphical models, any kind of trans-dimensional MCMC algorithm which is based on a discrete time Markov process (such as reversible jump algorithms by \citealp{wang2012efficient} and \citealp{lenkoski2013direct}) could suffer from high rejection rates, especially for high-dimensional graphs. However in our BDMCMC algorithm, moves between graphs are always accepted. In general, although our trans-dimensional MCMC algorithm has significant additional computing cost for birth and death rates, it has clear benefits over reversible jump style moves when graph structure learning in a non-hierarchical setting is of primary interest.

In Gaussian graphical models, Bayesian structure learning has several computational and methodological issues as the dimension grows: (1) convergence, (2) computation of the prior normalizing constant, and (3) sampling from the posterior distribution of the precision matrix. Our Bayesian approach efficiently eliminates these problems. For convergence, \citet{cappé2003reversible} demonstrate the strong similarity of reversible jump and continuous time methodologies by showing that, on appropriate rescaling of time, the reversible jump chain converges to a limiting continuous time birth-death process. In Section \ref{example 1} we show the fast convergence feature of our BDMCMC algorithm. For the second problem, in Subsection \ref{computing rates}, by using the ideas from \citet{wang2012efficient} and \citet{lenkoski2013direct} we show that the exchange algorithm circumvents the intractable normalizing constant. For the third problem, we used the exact sampler algorithm which was proposed by \citet{lenkoski2013direct}.

Our proposed method provides a flexible framework to handle the graph structure and it could be extended to different types of priors for the graph and precision matrix. In Subsection \ref{example: mammary data}, we illustrate how our proposed model can be integrated in types of graphical models, such as a multivariate time series graphical models. Although we have focused on normally distributed data, in general, we can extend our proposed method to other types of graphical models, such as log-linear models (see e.g. \citealp{dobra2011bayesian} and \citealp{lenkoski2011computational}), non-Gaussianity data by using copula transition (see e.g. \citealp{dobra2011copula}), or copula regression models (see e.g. \citealp{pitt2006efficient}). This will be considered in future work.

\section*{Appendix 1: Proof of theorem 1}
\label{proof theorem-bd}

Before we derive the detailed balance conditions for our BDMCMC algorithm we introduce some notation.

Assume the process is at state $(G, K)$, in which $G=(V,E)$ with precision matrix $K \in \mathbb{P}_{G}$. The process behavior is defined by the {\it birth rates} $\beta_{e}(K)$, the {\it death rates} $\delta_{e}(K)$, and the birth and death {\it transition kernels} $T^{G}_{\beta_{e}}(K;.)$ and $T^{G}_{\delta_{e}}(K;.)$. For each $e \in \Ehat$, $T^{G}_{\beta_{e}}(K;.)$ denotes the probability that the process jumps from state $(G, K)$ to a point in the new state $\cup_{K^{*} \in \mathbb{P}_{G^{+e} }}(G^{+e}, K^{*})$. Hence, if $ \mathcal{F} \subset \mathbb{P}_{G^{+e}}$ we have

\begin{eqnarray}
\label{kernel 1}
T_{\beta_e}^{G}(K;\mathcal{F}) = \frac{\beta_{e}(K)}{\beta(K)} \int_{k_{e} : K \cup k_{e} \in \mathcal{F}}{b_e(k_e ; K) dk_e}.
\end{eqnarray}
Likewise, for each $e \in E$, $T_{\delta_e}^{G}(K;.)$ denotes the probability that the process jumps from state $(G, K)$ to a point in the new state $\cup_{K^{*} \in \mathbb{P}_{G^{-e} }}(G^{-e}, K^{*})$. Therefore, if $\mathcal{F} \subset \mathbb{P}_{G^{-e}}$ we have

\begin{eqnarray*}
 T_{\delta_e}^{G}(K;\mathcal{F}) &=& \sum_{\eta \in E : K \setminus k_\eta \in \mathcal{F}}{\frac{\delta_{\eta}(K)}{\delta(K)}} \nonumber\\
                                 &=& \frac{\delta_{e}(K)}{\delta(K)} I(K^{-e} \in \mathcal{F}). 
\end{eqnarray*}

{\bf Detailed balance conditions.}  In our birth-death process, $P(K,G|\xb)$ satisfies detailed balance conditions if
\begin{eqnarray}
\label{balance1}
\int_{\mathcal{F}} \delta(K) dP(K,G|\xb)=\sum_{e \in E}{\int_{\mathbb{P}_{G^{-e}}} \beta(K^{-e}) T_{\beta_e}^{G}(K^{-e};\mathcal{F}) dP(K^{-e},G^{-e}|\xb)},
\end{eqnarray}
and
\begin{eqnarray}
\label{balance2}
\int_\mathcal{F} \beta(K) dP(K,G|\xb)=\sum_{e \in \Ehat}{\int_{\mathbb{P}_{G^{+e}}} \delta(K^{+e}) T_{\delta_e}^{G}(K^{+e};\mathcal{F}) dP(K^{+e},G^{+e}|\xb)},
\end{eqnarray}
where $\mathcal{F} \subset \mathbb{P}_{G}$.

The first expression says edges that enter the set $\mathcal{F}$ due to the deaths must be matched by edges that leave that set due to the births, and vice versa for the second part.

To prove the first part (\ref{balance1}), we have
\begin{eqnarray*}
LHS &=& \int_\mathcal{F} \delta(K) dP(G,K|\xb)                                                                                  \\
    &=& \int_{\mathbb{P}_{G}} I(K \in \mathcal{F}) \delta(K) dP(G,K|\xb)                                                        \\
    &=& \int_{\mathbb{P}_{G}} I(K \in \mathcal{F}) \sum_{e \in E}{\delta_{e}(K)} dP(G,K|\xb)                                    \\
    &=& \sum_{e \in E}{\int_{\mathbb{P}_{G}} I(K \in \mathcal{F}) \delta_{e}(K)} dP(G,K|\xb)                                    \\
    &=& \sum_{e \in E}{\int I(K \in \mathcal{F}) \delta_{e}(K)} P(G,K|\xb) \prod_{i=1}^{p} dk_{ii} \prod_{(i,j) \in E} dk_{ij}. \\
\end{eqnarray*}
For the RHS, by using (\ref{kernel 1}) we have
\begin{eqnarray*}
RHS &=& \sum_{e \in E}{\int_{\mathbb{P}_{G^{-e}}} \beta(K^{-e}) T_{\beta_e}^{G}(K^{-e};\mathcal{F}) dP(K^{-e},G^{-e}|\xb)}                             \\
        &=& \sum_{e \in E}{\int_{\mathbb{P}_{G^{-e}}} \beta_{e}(K^{-e}) \int_{k_{e} : K^{-e} \cup k_{e} \in \mathcal{F}}{b_e(k_e ; K^{-e}) dk_e} dP(K^{-e},G^{-e}|\xb)} \\
        &=& \sum_{e \in E}{\int_{\mathbb{P}_{G^{-e}}} \int_{k_{e}} I(K \in \mathcal{F}) \beta_{e}(K^{-e}) b_e(k_e ; K^{-e}) dk_e dP(K^{-e},G^{-e}|\xb)}              \\    
        &=& \sum_{e \in E}{\int I(K \in \mathcal{F}) \beta_{e}(K^{-e}) b_e(k_e ; K^{-e}) P(K^{-e},G^{-e}|\xb)}  \prod_{i=1}^{p} dk_{ii} \prod_{(i,j) \in E} dk_{ij}. \\
\end{eqnarray*}
By putting 
\begin{eqnarray*}
\delta_{e}(K) P(G,K|\xb) = \beta_{e}(K^{-e}) b_e(k_e ; K^{-e}) P(K^{-e},G^{-e}|\xb),
\end{eqnarray*}
we have LHS=RHS. Now, in the above equation 
\begin{eqnarray*}
P(G,K|\xb) = P(G,K \setminus (k_{ij},k_{jj})|\xb) P((k_{ij},k_{jj}) | K \setminus (k_{ij},k_{jj}),G, \xb),
\end{eqnarray*}
and
\begin{eqnarray*}
P(G^{-e},K^{-e}|\xb) = P(G^{-e},K^{-e} \setminus k_{jj}|\xb) P(k_{jj} | K^{-e} \setminus k_{jj},G^{-e}, \xb).
\end{eqnarray*}
We simply choose the proposed density for the new element $k_e = k_{ij}$ as follows:
\begin{eqnarray*}
 b_e(k_e ; K^{-e}) = \frac{P((k_{ij},k_{jj}) | K \setminus (k_{ij},k_{jj}),G, \xb)}{P(k_{jj} | K^{-e} \setminus k_{jj},G^{-e}, \xb)}.
\end{eqnarray*}
Therefore, we reach the expression in Theorem \ref{BD theorem}. The proof for the second part (\ref{balance2}) is the same.

\section*{Appendix 2: Proposition}
\label{appendix:proposition}

Let A be a $2 \times 2 $ random matrix with Wishart distribution $W(b, D)$ as below
\begin{eqnarray*}
P(A) = \frac{1}{I(b,D)} |A|^{(b-2)/2} \exp \left\{ - \frac{1}{2} \mbox{tr}(DA) \right\},
\end{eqnarray*}
where 
\begin{eqnarray*}
A =
\begin{bmatrix}
 a_{11}  & a_{12}    \\
 a_{12}  & a_{22}    \\
\end{bmatrix},
\qquad
D =
\begin{bmatrix}
 d_{11}  & d_{12}    \\
 d_{12}  & d_{22}    \\
\end{bmatrix}.
\end{eqnarray*}
Then

(i) $a_{11} \sim W(b+1, D_{11.2})$ where $D_{11.2} = d_{11} - d_{22}^{-1} d_{21}^{2}$,

\vspace{2 mm} 

(ii) 
\begin{eqnarray*}
P(a_{12}, a_{22} | a_{11}) &=& \frac{P(A)}{P(a_{11})}                                        \\
    &=& \frac{1}{J(b,D,a_{11})} |A|^{(b-2)/2} \exp \left\{  - \frac{1}{2} \mbox{tr}(DA) \right\},   \\
\end{eqnarray*}
where 
\begin{eqnarray*}
J(b,D,a_{11}) = \left( \frac{2 \pi}{d_{22}} \right)^ {\frac{1}{2}} I(b,d_{22}) a_{11}^{\frac{(b-1)}{2}} \exp \left\{  - \frac{1}{2} D_{11.2} a_{11} \right\}.
\end{eqnarray*}

{\bf Proof.} For proof of part (i), see \citet[Theorem 3.2.10]{muirhead1982aspects}. The result for part (ii) is immediate by using part (i).

\bibliographystyle{ba}
\bibliography{references}

\begin{acknowledgement}
The authors wish to thank two referees and an associate editor for their suggestions which led to significant improvement of this paper. The authors are also grateful to Alex Lenkoski for the code for the direct sampler algorithm from the G-Wishart distribution and to Hao Wang for his suggestions. 
\end{acknowledgement}

\end{document}